\documentclass[%
reprint,
superscriptaddress,
groupedaddress,
 amsmath,amssymb,
 aps,
prb,
floatfix,
]{revtex4-1}

\usepackage{graphicx}% Include figure files
\usepackage{dcolumn}% Align table columns on decimal point
\usepackage{bm}% bold math
\usepackage{graphicx}
\usepackage{placeins}
\begin{document}

\title{Effect of random pinning on nonlinear dynamics and dissipation of a vortex driven by a strong microwave current.}

\author{W. P. M. R. Pathirana and A. Gurevich.} 
\affiliation{
Department of Physics and Center for Accelerator Science, Old Dominion University, Norfolk, Virginia, USA 
}
	
%\maketitle

%
\begin{abstract}
	
We report numerical simulations of a trapped elastic vortex driven by a strong ac magnetic field $H(t)=H\sin\omega t$ parallel to the surface of a superconducting film. The surface resistance and the power dissipated by an oscillating vortex perpendicular to the film surface were calculated as functions of $H$ and $\omega$ for different spatial distributions, densities and strengths of pinning centers, including bulk pinning, surface pinning and cluster pinning. Our simulations were performed for both the Bardeen-Stephen viscous vortex drag and the Larkin-Ovchinnikov (LO) drag coefficient $\eta(v)$ decreasing with the vortex velocity $v$. The local residual surface resistance $R_i(H)$ calculated for different statistical realizations of the pinning potential exhibits strong mesoscopic fluctuations caused by local depinning jumps of a vortex segment as $H$ increases, but the global surface resistance $\bar{R}_i(H)$ obtained by averaging $R_i(H)$ over different pin configurations increases smoothly with the field amplitude at small $H$ and levels off at higher fields. For strong pinning, the LO decrease of $\eta(v)$ with $v$ can result in a nonmonotonic field dependence of $R_i(H)$ which decreases with $H$ at higher fields, but cause a runaway instability of the vortex in a thick film for weak pinning. It is shown that overheating of a single moving vortex can produce the LO-like velocity dependence of $\eta(v)$, but can mask the decrease of the surface resistance with $H$ at a higher density of trapped vortices. 

\end{abstract}
 
 \maketitle

\section{Introduction}

Dynamics of vortices driven by electric currents and pinning of vortices by materials defects determine electromagnetic properties of superconductors in a magnetic field. The ability of type-II superconductors to carry weakly-dissipative current densities $J$ up to a critical current density $J_c$  is crucial for many applications ~\cite{mag1,mag2,mag3,ce}. Mechanisms by which the vortex matter is pinned by materials defects have attracted renewed attention since the discovery of high-$T_c$ cuprates for which $J_c$ is controlled by complex interplay of pinning, interaction of elastic vortices, and thermal fluctuations ~\cite{blatter,ehb}. Advances in optimization of pinning nanostructures in superconductors have pushed $J_c$ up to $10-30\%$ of the depairing current density $J_d$ at which current breaks Cooper pairs ~\cite{mag1,mag2,jc1,jc2,jc3,jc4,jc5}.   

The physics of depinning of long vortices by a uniform dc current has been well established both for weak  collective pinning ~\cite{blatter,col} and strong pinning ~\cite{ehb,sp1,sp2,sp3,sp4,sp5,sp6}. Usually pinning potential is assumed random, although possibilities of enhancing $J_c$ by quasi-periodic, conformal, graded or hyper uniform pinning have been considered  ~\cite{grad1,grad2,grad3,grad4,grad5}. In the case of collective pinning of a long vortex $J_c$ is a self-averaging quantity which remains the same for any position of a vortex  in a statistically-uniform pinning potential. This property is also characteristic of vortices parallel to the surface subject to an ac magnetic field ~\cite{ce,cam1,cam2,cam3,cam4} for which a low-field surface impedance ~ \cite{ce,blatter,ehb,coffey} and a hysteretic electromagnetic response at strong ac field ~\cite{cam3,cam4} have been thoroughly investigated in the literature. Here $J_c$ remains a self-averaging characteristic because long vortices parallel to the surface are pinned by multiple defects and are driven by a uniform ac current.  

\begin{figure}[!htb]
	\centering
	\includegraphics[scale=0.25]{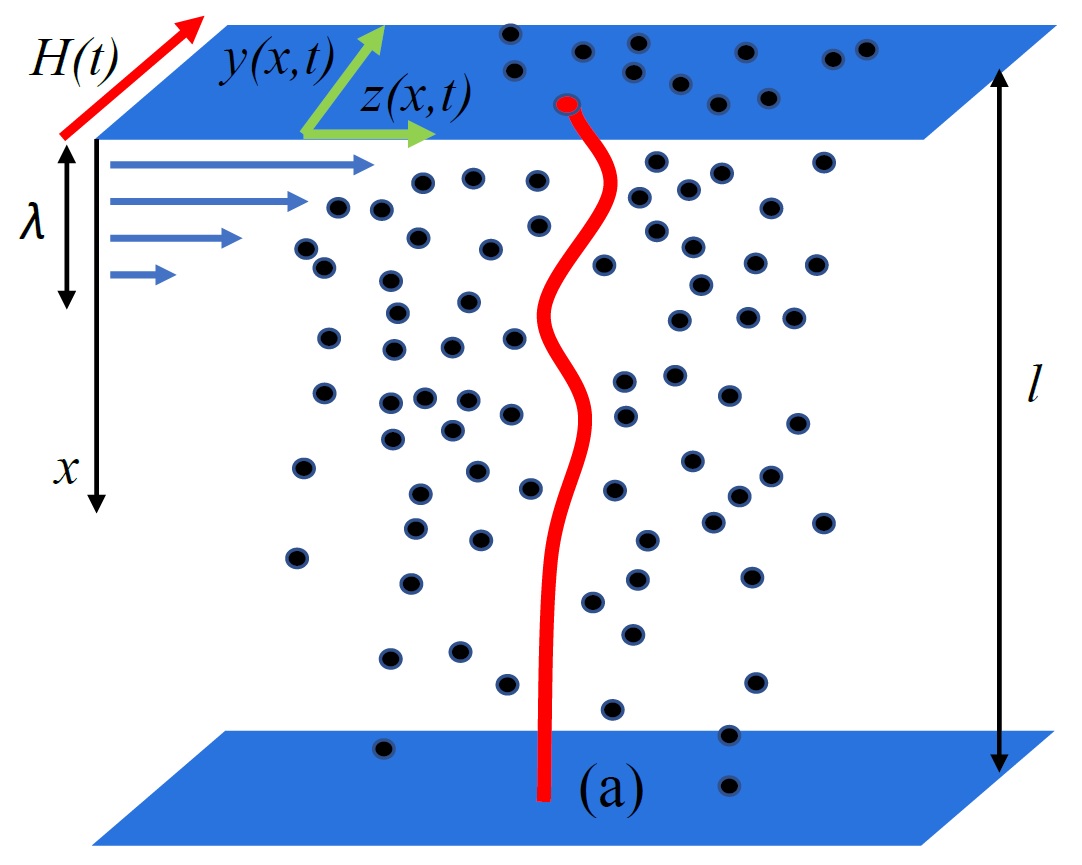}
	\includegraphics[scale=0.5]{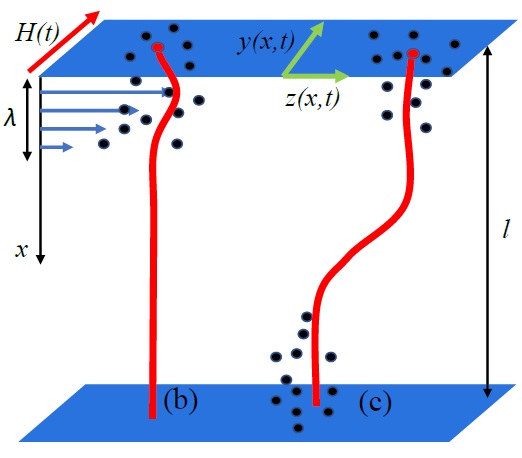}
	\caption{\label{fig:Fig1} A trapped vortex driven by the rf surface current for different distributions of pinning centers shown by black dots: (a) bulk pinning (b) surface pinning (c) cluster pinning. Green arrows show vortex tip displacement on the YZ plane.} 
\end{figure}

A different situation occurs for sparse vortices perpendicular to the surface. Here vortices are driven by the Meissner current flowing in a thin layer at the surface so the Lorentz force is only applied to a tip of a vortex, as shown in Fig. \ref{fig:Fig1}. The resulting bending distortions of an elastic vortex extends over the Campbell length  ~\cite{ce,cam1,cam2,cam3,cam4}, so that a vibrating vortex segment interacts only with a few pins, while the rest of a long vortex does not move. In this case the response of the vortex becomes dependent on its position in a particular configuration of pinning centers. Shown in Fig. 1 are representative cases of bulk pinning, pins segregated randomly at the surface and clusters of pins.  The global electromagnetic response is a sum of responses of individual vortices moving in their respective pinning potentials which can fluctuate strongly along the surface. Similar pinning fluctuations cause local variation of $J_c$ of perpendicular vortices in thin films ~\cite{ehb,brandt1,ag_pin} and can also play a role in bulk pinning ~\cite{geshk2}.       

Power dissipated by sparse vortices driven by radio-frequency (rf) magnetic field $H(t)=H\sin\omega t$ is an important characteristic of superconducting structures with extremely high quality factors $Q$, particularly resonator cavities for particle accelerators ~\cite{Padam_book,gurevich2012}, micro cavities and thin film resonators ~\cite{microcav}. For Nb resonant cavities, the quality factor $Q$ can reach $10^{10}-10^{11}$ at 1-2 K and 1 GHz in the  Meissner state. The rf power per unit area $P(H)=R_s(H)H^2/2$ is determined by the surface resistance $R_s=R_{BCS}+R_i$ which contains the quasiparticle BCS resistance $R_{BCS}\propto\omega^2\exp(-\Delta/T)$~  \cite{tinkham} and a weakly-temperature dependent residual resistance $R_i$. The main contribution to $R_i$, which can significantly exceed $R_{BCS}$, comes from trapped vortices generated during slow cool down through $T_c$ ~\cite{tf1,tf2,tf3,tf4,tf5,jlab_20,fnal_20}. As a result, stray fields of a few $\%$ of the Earth's magnetic field can produce vortices trapped by material defects and give rise to rf hotspots ~\cite{tf2}. 

Low-field rf losses of pinned vortices have been calculated by many authors ~\cite{ehb,gc,tf2,ag_sust,jlab_20}. Nonlinear quasi-static electromagnetic response of perpendicular vortices has been addressed for both weak collective pinning  ~\cite{liarte} and strong pinning  ~\cite{cam3,cam4}.   A variety of field dependencies of $R_s(H)$ have been observed, including a quasi-linear increase of $ R_s(H) $ with $H$ at low field and saturation at higher field ~\cite{jlab_20,liarte} or descending $R_s(H)$ ~\cite{gigi}. Yet the behavior of $R_s(H)$ controlled by the nonlinear dynamics of a flexible vortex driven by a strong surface rf current through pinning centers is poorly understood. 

For fast vortices driven by Meissner current with $J\gg J_c$, the velocity of the vortex $v$ is mainly determined by a balance of the Lorentz force $F_L=\phi_0J$ and the viscous drag force, $F_d=\eta(v)v$, where the vortex drag coefficient $\eta$ can essentially depend on $v$. Here $v\simeq \phi_0J/\eta(v)$ can exceed the pairbreaking superfluid velocity of the condensate $v_d=\Delta/p_F$ at $J<J_d$, where $\phi_0$ is the flux quantum, $\Delta$ is the superconducting gap and $p_F$ is the Fermi momentum ($v_d\simeq 1$ km/s for Nb). Vortices moving faster than the superflow which drives them have been observed by scanning SQUID on tip microscopy on Pb films in which $v$ can exceed $v_d$ by two orders of magnitude ~\cite{embon}. Such high velocities may result from the Larkin-Ovchinnikov (LO) mechanism in which $\eta(v)$ decreases with $v$ as the moving vortex core becomes depleted of nonequilibrium quasiparticles lagging behind ~\cite{LO}.  The LO theory  predicts a nonmonotonic velocity dependence of the drag force $F_d=\eta(v)v$ which cannot balance the Lorentz force if $v$ exceeds a critical value $v_0$. The LO instability has been observed by dc transport measurements on many superconductors ~\cite{lo1,lo2,lo3,lo4,lo5,lo6,lo7,lo8,lo9,lo10,lo11} with typical values of $v_0\sim 0.1-1$ km/s near $T_c$, the LO instability at low $T$ being  masked by heating effects. Heating is weaker if the vortex is driven by Meissner rf current, in which case the LO velocity dependence of $\eta(v)$ can produce $R_i(H)$ decreasing with $H$ ~\cite{Manula}. Other mechanisms of the velocity-dependent $\eta(v)$ and instability of flux flow can result from overheating of moving vortices ~\cite{shklovsk,kunchur,gc} or elongation of the vortex core along the direction of motion at $v>v_d$ revealed by simulations of the time-dependent Ginzburg-Landau (TDGL) equations ~\cite{embon,tdgl1,tdgl2}. 

Addressing the mechanisms of dissipation of vortices in the case of mesoscopic pinning and a velocity-dependent $\eta(v)$ requires computer simulations of nonlinear dynamics of an elastic vortex driven by a strong Meissner current. Such simulations are reported in this paper in which we calculated the field and frequency dependencies of $R_i(H,\omega)$ for realistic pinning structures shown in Fig. \ref{fig:Fig1}. In particular, we calculated the effect of the LO velocity dependence of $\eta(v)$ on $R_i(H,\omega)$ in a film with many pinning centers, extending our previous results for a vortex pinned by a single material defect ~\cite{Manula}. We also addressed the overheating of a single moving vortex and its effect on the nonlinear vortex drag, as well as  the effect of overheating of sparse vortices on the field and frequency dependencies of the global surface resistance.  

The paper is organized as follows. Section II introduces the main nonlinear dynamic equations for a trapped curvilinear vortex and defines key control parameters. In Sec. III we present numerical simulations of a vortex at low fields and calculate the field and frequency dependencies of $R_i(H,\omega)$ averaged over statistical realizations of random pinning potential. Section IV contains the numerical results for a vortex driven by strong fields, including the issue of a depinning rf field and the effect of the LO velocity-dependent vortex drag on $R_i(H,\omega)$ in the presence of pinning. Section V addresses the overheating effects caused by driven vortices, particularly the LO-like $\eta(v)$ produced by overheating of a single vortex.  Section VI contains a discussion of our results.    

\section{Dynamic equations.} \label{DE}

For the geometry shown in Fig. \ref{fig:Fig1}, the dynamic equation for the coordinates ${\bf u}=[u_y(x,t),u_z(x,t)]$ of the vortex moving in the $yz$ plane is given by:
\begin{gather} 
M \frac{\partial^2 {\bf u}}{\partial t^2}+\eta \frac{\partial {\bf u}}{\partial t}=\epsilon\frac{\partial^2{\bf u}}{\partial x^2}-\nabla U(x,{\bf u})+\hat{y}f_L(x,t),
\label{eq1} \\
f_L(x,t)=(\phi_0H/\lambda)e^{-x/\lambda}\sin\omega t,
\label{eq2}
\end{gather}
where $H$ is the amplitude of the applied magnetic field $H\sin\omega t$ with the frequency $f=\omega/2\pi$, $\lambda$ is the London penetration depth, $M$ is the vortex mass per unit length, $\epsilon=\phi_0^2(\ln\kappa+0.5)/4\pi\mu_0\lambda^2$ is the vortex line energy, $ \kappa=\lambda/\xi $ is the GL parameter, $\xi$ is the coherence length, and $\eta(v)$ is a vortex drag coefficient.

Equations (\ref{eq1}) and (\ref{eq2}) represent a balance of local forces acting on a curvilinear vortex: the inertial and drag forces in the left hand side are balanced by the elastic, pinning and Lorentz forces in the right hand side.  It is assumed that:  1. $H$ is well below the superheating field  ~\cite{hs1,hs2,hs3,hs4} so that the London model is applicable. 2. The Magnus force causing a small Hall angle ~\cite{v1,v2,v3} is negligible. 3. The low frequency rf field ($\hbar\omega\ll \Delta$) does not produce quasiparticles, and the quasi-static London equations are applicable ~\cite{tinkham}. 4. Bending distortions of the vortex are small and the linear elasticity theory \cite{blatter,ehb} can be used. The effect of nonlinear elasticity was addressed recently ~\cite{Manula}.  5. The elastic nonlocality ~\cite{blatter,ehb} is neglected. The effect of nonlocality of $\epsilon$ on the low-field vortex losses was addressed previously~ \cite{tf1}.

We consider here the core pinning of vortices ~\cite{ce,blatter,ehb} represented by a sum of pinning centers modeled by the Lorentzian functions ~\cite{Embon_2}:
\begin{equation} 
\!U(x,{\bf u})=-\sum_{n=1}^{N}\frac{U_n}{1+[(x-x_n)^2+|{\bf u}-{\bf r}_n|^2]/\xi^2}.
\label{eq3}
\end{equation}
Here, $x_n, {\bf r}_n=(y_n,z_n)$ are the coordinates of the n-th pinning center and $U_n$ is determined by the gain in the condensation energy in the vortex core at the pin ~\cite{ce,blatter,ehb}.

At high vortex velocities $\eta(v)$ can depend on $v$. For instance, in the LO model $\eta(v)$ is given by  ~\cite{LO}:
\begin{gather} 
\eta=\frac{\eta_0 }{1+v^2/v_0^2},
\label{LO}\\
v_0^2=\frac{D\sqrt{14\zeta(3)}}{\pi\tau_\epsilon}\left(1-\frac{T}{T_c}\right)^{1/2}.
\label{vol}
\end{gather}
Here $\eta_0=\phi_0^2/2\pi \xi^2\rho_n$ is the Bardeen-Stephen drag coefficient ~\cite{tinkham}, $D$ is the electron diffusivity, $\tau_\epsilon$ is the energy relaxation time of quasiparticles, and $\zeta(3)\approx 1.202$. A similar $\eta(v)$ can also result from overheating of moving vortices  ~\cite{shklovsk,kunchur,gc}. If the energy relaxation is limited by electron-phonon collisions, then
\begin{equation}
\tau_\epsilon=\frac{2\hbar c_s^2p_F^2}{7\pi\zeta(3)\lambda_{ep}(k_BT)^3},\qquad T\approx T_c
\label{taup}
\end{equation}
where $c_s$ is the speed of longitudinal sound, $p_F=\hbar(3\pi^2n)^{1/3}$, $n$ is the carrier density, and $\lambda_{ep}$ is a dimensionless electron-phonon coupling constant  ~\cite{kopnin}. 

The LO model predicts a nonmonotonic velocity dependence of the drag force $F_d=\eta(v)v$ which can balance the Lorentz force $F_L=\phi_0J$ only if $v<v_0$ and $F_L<\eta_0v_0/2$.  Jumps on voltage-current characteristics caused by the LO instability have been observed on many superconductors ~\cite{lo1,lo2,lo3,lo4,lo5,lo6,lo7,lo8,lo9,lo10,lo11} with $v_0\sim 0.1-1$ km/s near $T_c$. These experiments have shown that as $T$ decreases, $v_0(T)$ first increases near $T_c$ and then decreases at lower temperatures ~\cite{lo3,lo9}, consistent with Eqs. (\ref{vol})-(\ref{taup}). 

Combining Eqs. (\ref{eq1})-(\ref{LO}), we obtain the following nonlinear equations for the dimensionless coordinates of the vortex $u_y(x,t)=u_y/\lambda$ and $u_z(x,t)=u_z/\lambda$:
\begin{gather} 
\mu\ddot{u}_y+\frac{\gamma\dot{u}_y}{1+\alpha (\dot{u}_y^2+\dot{u}_z^2)}=
\nonumber \\
u_y''-\sum_{n=1}^{N}A_n(x,{\bf u})(u_y-y_n)+\beta_t e^{-x},
\label{dyneq1}\\
\!\!\!\mu\ddot{u}_z+\frac{\gamma\dot{u}_z}{1+\alpha (\dot{u}_y^2+\dot{u}_z^2)}=u_z''-\!\sum_{n=1}^{N}A_n(x,{\bf u})(u_z-z_n),
\label{dyneq2} \\
u_y'(0,t)=u_z'(0,t)=u_y'(l,t)=u_z'(l,t)=0.
\label{bc0}
\end{gather}
Here the prime and the overdot imply differentiation over the dimensionless coordinate $x=x/\lambda$ and time $t = tf$, respectively,  and:
\begin{gather}
\gamma=f/f_0,  \qquad f_0=H_{c1}\rho_n/{H_{c2}\lambda^2 \mu_0},  
\label{gamma}\\
\beta_t=\beta \sin(2\pi t), \qquad \beta=H/H_{c1},
\label{beta}\\
\alpha=\alpha_0\gamma^2, \qquad \alpha_0=(\lambda f_0/v_0)^2,
\label{alpha}\\
\mu=\mu_1 \gamma^2, \qquad \mu_1=\lambda^2f_0^2 M/\phi_0 H_{c1},
\label{mu}\\
A_n=\frac{\zeta_n}{\left[1+\kappa^2(x-x_n)^2+\kappa^2|{\bf u}-{\bf r}_n|^2\right]^2},
\label{alphan}\\
\zeta_n=2\kappa^2U_n/\epsilon,
\end{gather}
where  $H_{c1}=(\phi_0/4\pi\mu_0\lambda^2)(\ln\kappa+0.5)$ and $H_{c2}=\phi_0/2\pi\mu_0\xi^2$ are the lower and upper critical fields, respectively. The vortex mass $M_s\simeq 2p_F/\hbar\pi^3$ results from quasiparticles in the vortex core ~\cite{suhl}, but other mechanisms can produce $M$ much larger than $M_s$ ~\cite{vm1,vm2,vm3,vm4}. For instance, $M\sim 10^2M_s$  was observed in Nb near $T_c$ ~\cite{golubchik}.  

The amplitude $U_n$ in Eq. (\ref{eq3}) determines the elementary pinning energy $u_p=\pi\xi U_n$ and the pinning parameter $\zeta_n=2\kappa^2 u_p/\pi\epsilon\xi$. For a dielectric precipitate of radius $r_0<\xi$, we have $u_p\sim B_c^2r_0^3/\mu_0$ and $\zeta_n\sim (r_0/\xi)^3\kappa^2$, where $B_c=\phi_0/2^{3/2}\pi\xi\lambda$ is the GL thermodynamic critical field ~\cite{ce}. For a single impurity with a scattering cross-section $\sigma_i$, we have  $u_p\sim B_c^2\sigma_i\xi/\mu_0$ ~\cite{thun} and $\zeta_n\sim \sigma_i\kappa^2/\xi^2$. In both cases $\zeta_n$ can be larger than $1$ if $\kappa\gg 1$. Equations (\ref{dyneq1})-(\ref{dyneq2}) describe an elastic vortex interacting with pinning centers. The case in which Eq. (\ref{dyneq2}) is disregarded ~\cite{fnal_20} is more relevant to a vortex interacting with perpendicular columnar defects. 

For Nb with $\rho_n\approx3$ n$\Omega$m, $\lambda=80$ nm, $\xi=20$ nm $\kappa=4$, $v_0=0.1$ km/s, $k_F=1.2\times 10^{10}$ m$^{-1}$ (see, Ref. \onlinecite{ashcroft}), and $M=80M_s=5.6\times 10^{-20}$ kg/m, we have $f_0\simeq 22$ GHz, $\alpha_0\simeq 309$, and $\mu_1\simeq 0.0022$, so that $\gamma\simeq 0.045$, $\mu\simeq 4.5\cdot 10^{-6}$, and $\alpha\simeq 0.64$ at $f=1$ GHz. For  Nb\textsubscript{3}Sn with $\rho_n\approx 1$  
$\mu\Omega$m, $\lambda=111$ nm, $\xi=4.2$ nm, $\kappa=26.4$ ~\cite{liarte}, $v_0=0.1$ km/s, $k_F=6.6\times 10^{9}$ m$^{-1}$, and $M=80M_s=3.1\cdot 10^{-20}$ kg/m, we have $f_0\simeq 175$ GHz, $\alpha_0\simeq 3.7\times 10^{4}$, and $\mu_1\simeq 0.14$ so that $\gamma\simeq 0.006$, $\mu\simeq 4.6\cdot 10^{-6}$, and $\alpha\simeq 1.2$ at $f=1$ GHz. 

Another key parameter is a complex penetration length $L_\omega$ of bending distortions induced by the surface rf Meissner current along the vortex  ~\cite{ag_sust}:                      
\begin{equation}
L_\omega=\sqrt{\frac{\epsilon}{k_L+i\eta\omega}},
\label{Lom}
\end{equation}
where $k_L\sim \phi_0J_c/\xi$ is the Labusch pinning spring constant~\cite{ce}. At $\omega\eta\ll k_L$  Eq. (\ref{Lom})  reduces to the Campbell penetration depth \cite{cam1,cam2,cam3,cam4} or the Larkin pinning length $L_c \sim \xi \sqrt{J_d/J_c}$ in the collective pinning theory ~\cite{col,blatter,ehb}. At high frequencies, $\omega\eta\gg k_L$, Eq. (\ref{Lom}) yields the elastic skin depth $L_\omega \rightarrow [\epsilon/\eta\omega]^{1/2}$. For Nb$_3$Sn, we have $L_\omega\simeq 5.15\lambda=572$ nm at 1 GHz, and $L_\omega\simeq 52\lambda=5.7\,\mu$m at 10 MHz, so the rf distortions along the vortex extend well beyond the field penetration depth.

To compare $L_\omega$ with the quasiparticle diffusion length $L_d\sim \sqrt{D\tau_\epsilon}$, 
we present $L_d$ and $v_0$ in the form: 
\begin{equation}
L_d\sim\xi_0\sqrt{\frac{\Delta_0\tau_\epsilon}{\hbar}},\quad v_0=1.6\xi_0\sqrt{\frac{\Delta_0}{\hbar\tau_\epsilon}}\left(1-\frac{T}{T_c}\right)^{1/4},
\label{diff}
\end{equation}
where $\xi_0=\sqrt{\hbar D/2\Delta_0}$ is a coherence length in the dirty limit and $\Delta_0$ is a superconducting gap at $T=0$. For an alloyed Nb with $\xi_0=10$ nm, $n=7\times 10^{28}$ m$^{-3}$ extracted from the Hall measurements ~\cite {hall1}, $c_s=3.48$ km/s ~\cite{ashcroft}, $\lambda_{ep}=1$, $T_c=9$K, $\Delta_0=1.8k_BT_c$, we obtain $\tau_c=\tau_\epsilon(T_c)\simeq 9\times 10^{-11}$s. Then Eq. (\ref{diff}) gives $\tau_\epsilon\simeq 8.8\times 10^{-10}$s, $L_d\simeq 43\xi_0$, $v_0\simeq 0.68$ km/s at $4.2$ K, and $\tau_\epsilon \simeq 8.2\times 10^{-9}$s, $L_d\simeq 132\xi_0$, $v_0\simeq 0.24$ km/s at $2$ K. For Nb$_3$Sn with $\xi_0=3$nm, $\Delta_0=1.9k_BT_c$, $T_c=18$K ~\cite{hc2},  $c_s=4.7$ km/s ~ \cite{cs}, $n=1.77\times 10^{28}$ m$^{-3}$ ~\cite{hall2},  $\lambda_{ep}=1.7$, we obtain $\tau_c\simeq 4.8\times 10^{-12}$s. Hence, $\tau_\epsilon\simeq 3.8\times 10^{-10}$s, $L_d\simeq 42\xi_0$, $v_0\simeq 0.48$ km/s at $4.2$ K.

The above rough estimates indicate that, there is a broad range of $T$ and $\omega$, where $\omega t_\epsilon < 1$ and $L_d \ll L_\omega$. In this case the diffusive cloud of nonequilibrium quasiparticles follows adiabatically behind the vortex core curved weakly over the length $\sim L_d$ and Eq. (\ref{LO}) may be applicable for a curvilinear oscillating vortex as well. This was assumed in the simulations described in Sec. \ref{Larkin}, where we also show that $L_\omega$ increases significantly as the vortex velocity increases and $\eta(v)$ decreases. 

Equation (\ref{LO}) was obtained for a single vortex ~\cite{LO}, but in dc transport measurements on film bridges in a perpendicular field $B_0$, the LO velocity is extracted by fitting the data with $v_0=v_1+AB_0^{-1/2}$, where $v_1$ and $A$ are constants ~\cite{lo3}. This form of $v_0(B_0)$ phenomenologically takes into account a reduction of $v_0$ due to overlapping clouds of nonequilibrium quasiparticle from neighboring vortices at $B_0\gtrsim B_T\sim \phi_0/L_d^2$. Extracting $v_0$ at $B_0\ll B_T$ from magneto-transport measurements is ambiguous because both the dc transport current density $J(y)$ and the vortex density can be highly inhomogeneous across the film, causing a vortex dome in the middle of the bridge, peaks in $J(x)$ at the film edge and a geometrical barrier for the vortex entry~ \cite{embon}. The LO single-vortex limit could be probed in the geometry shown in Fig.\ref{fig:Fig1} where the Meissner current is uniform over the surface. We assume that $v_0(B_0)$ could depend on the local density of trapped vortices in hotspots if $B_0>B_T$ but the vortex spacing is larger than $\lambda$ so vortices do not interact and the clouds of nonequilibrium quasiparticles affect weakly the line tension and pinning energies in the force balance Eqs. (\ref{eq1}) and (\ref{eq2}).

Overheating ~\cite{shklovsk,kunchur,gc} and the vortex core stretching at high velocities ~\cite{embon,tdgl1,tdgl2} can also produce the LO-like $\eta(v)$ at $T\simeq T_c$. However, no microscopic theory of $\eta(v)$ at low temperatures has been developed, and the TDGL equations are not applicable at $T\ll T_c$. Yet the physical mechanisms behind the LO form of $\eta(v)$ such as the depletion of nonequilibrium quasiparticles in the moving vortex core and overheating described in Sec. \ref{Heat} would operate at $T\ll T_c$ as well. We use the representative Eq. (\ref{LO}) in our numerical simulations to reveal distinctive qualitative features of the surface resistance resulting from the interplay of pinning and the descending velocity dependence of $\eta(v)$. 

The surface resistance $R_i$ defines the power of rf losses $P=R_iH^2/2=\phi_0\sum_k\int \langle J(x,t) \partial_t u_y^{k}(x,t) dx\rangle$ per unit area from all vortices, where $u_y^{k}(x,t)$ describes the $k-$th vortex and $\langle ... \rangle$ means time averaging (see Appendix A).  For small density of vortices, it is convenient to define a mean dimensionless power $p=P/P_0$ and the surface resistance $r_i$ per vortex:  
\begin{gather} 
p=\frac{\gamma}{N_v} \sum_{k=1}^{N_v}\int_{0}^{1}dt\int_{0}^{l}\beta_{t}e^{-x}\dot{u}_y^k(x,t)dx,
\label{p} \\
r_i(\beta)=2p(\beta)/\beta^2,
\label{rs}
\end{gather}
where $P_0=\lambda f_0 \epsilon$, $N_v$ is the number of vortices and $R_i=P_0r_i n_\square/H_{c1}^2$. Here $n_\square=B_0/\phi_0$ is a vortex areal density proportional to a small induction $B_0\ll B_{c1}$.  Using here $f_0$ from Eq. (\ref{gamma}) and $\epsilon=\phi_0H_{c1}$, we obtain:

\begin{equation}
R_i=\frac{\rho_nB_0}{\lambda B_{c2}} r_i.
\label{rii}
\end{equation}

\section{Numerical Results} \label{Results}

We solved Eqs. (\ref{dyneq1})-(\ref{bc0}) numerically using COMSOL \cite{comsol}.  In our simulations a straight vortex was initially put in a particular pinning potential and after ${\bf u}(x,t)$ relaxes to a stable shape, the rf field was turned on. Then we run the program until ${\bf u}(x,t)$ reaches steady-state oscillations after a transient period $\delta t\lesssim 90/f$ and use this solution to calculate $R_i$. This section presents the results for $R_i(H,\omega)$ at $H\lesssim H_{c1}$ and $v\ll v_0$ for which $\eta_0$ is independent of $v$, and the vortex mass is neglected.  

\subsection{Bulk pinning} \label{bulk}

For bulk pinning shown in Fig. \ref{fig:Fig1}a, $N$ identical pins were distributed randomly in a $l_y\times l_z\times l$ box, where the length $l$ along the vortex was varied from $5\lambda$ to $50\lambda$. Equations (\ref{dyneq1})-(\ref{bc0}) were solved for different $N=10, 50, 100$, making sure that $l_y$ and $l_z$ are adjusted in such a way that the vortex always remains within the box during the rf period. The results are expressed in terms of a mean pin density $n_i=N/ll_yl_z$.

\begin{figure}[h]
	\centering
	\includegraphics[width=\columnwidth]{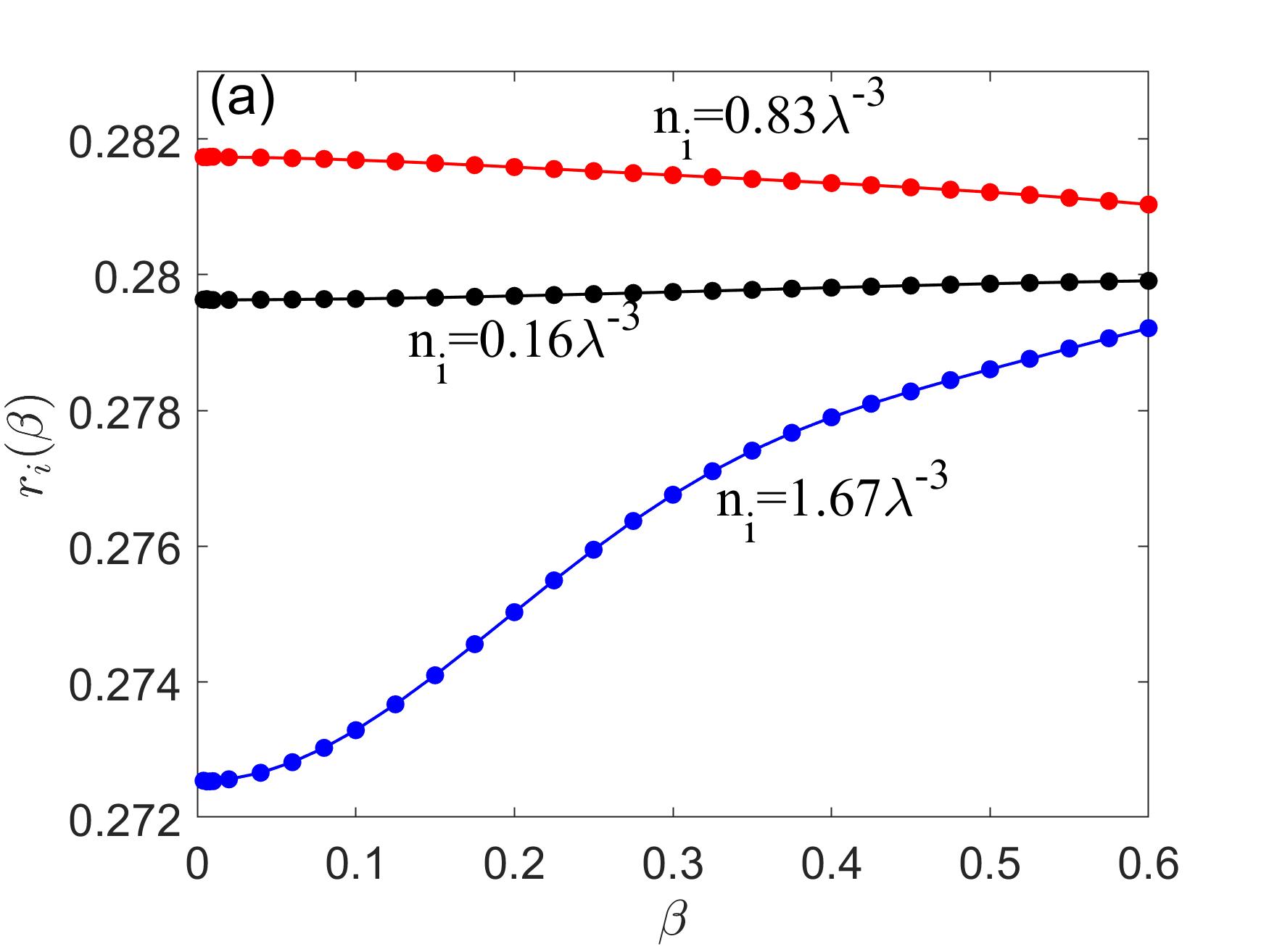}\\
	\includegraphics[width=\columnwidth]{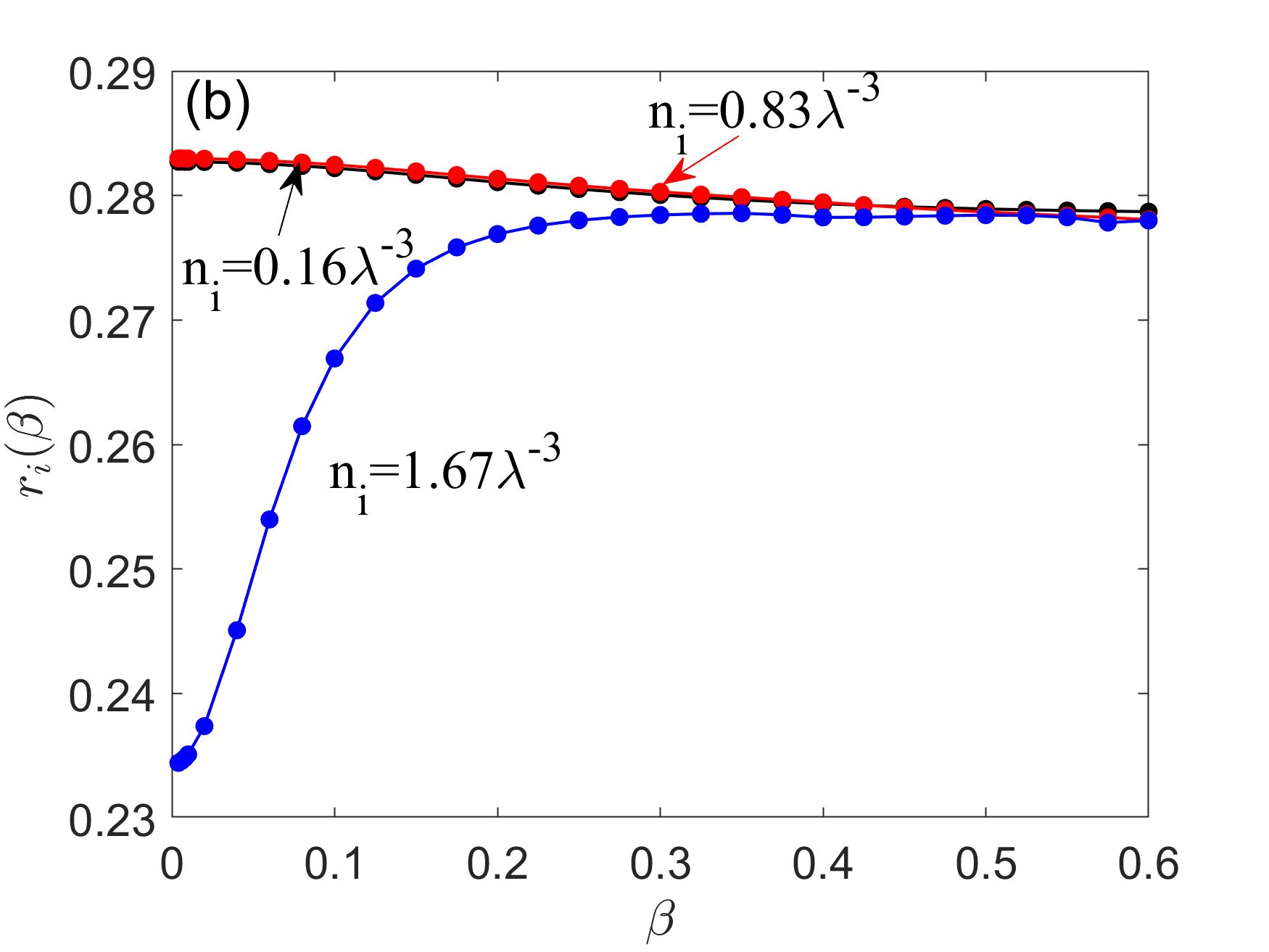}\\ 
	\caption{ Field dependence of $r_i(\beta) $ at $l/\lambda=10$, $\gamma=0.04$, $n_i=0.16 \lambda^{-3}, 0.83 \lambda^{-3}, 1.67 \lambda^{-3}$, and: (a) $\kappa=2$, $\zeta_n=0.04$; (b) $\kappa=10$, $\zeta_n=1$.} 
	\label{fig:Fig2}
\end{figure} 

Shown in Fig. \ref{fig:Fig2} are the dependencies of the surface resistance $r_i(\beta)$ on the field amplitude $\beta=H/H_{c1}$ calculated for different pin densities $n_i$ at $\kappa=2$ and $\kappa=10$.  Here $r_i$ is nearly independent of $H$ for smaller pin densities but develops a field dependence at $ n_i=1.67\lambda^{-3}$. Curiously, $ r_i(\beta)$ at $n_i=0.16\lambda^{-3}$ is slightly smaller than $r_i(\beta)$ at $n_i=0.83\lambda^{-3}$, which reflects the effect of pinning fluctuations on $r_i$. At $\beta\gtrsim 0.2-0.4$ all $r_i(\beta)$ curves calculated for different $n_i$ approach a field-independent limit $r_i\approx 0.28$ controlled by the vortex drag in which the effect of pinning on $r_i$ is weak as $J\gg J_c$.  At $n_i=1.67\lambda^{-3}$ pinning mitigates vortex oscillations and reduces $r_i(\beta)$ at low fields $\beta\lesssim 0.2-0.4$.  

\begin{figure}[h]
	\centering
	\includegraphics[width=\columnwidth]{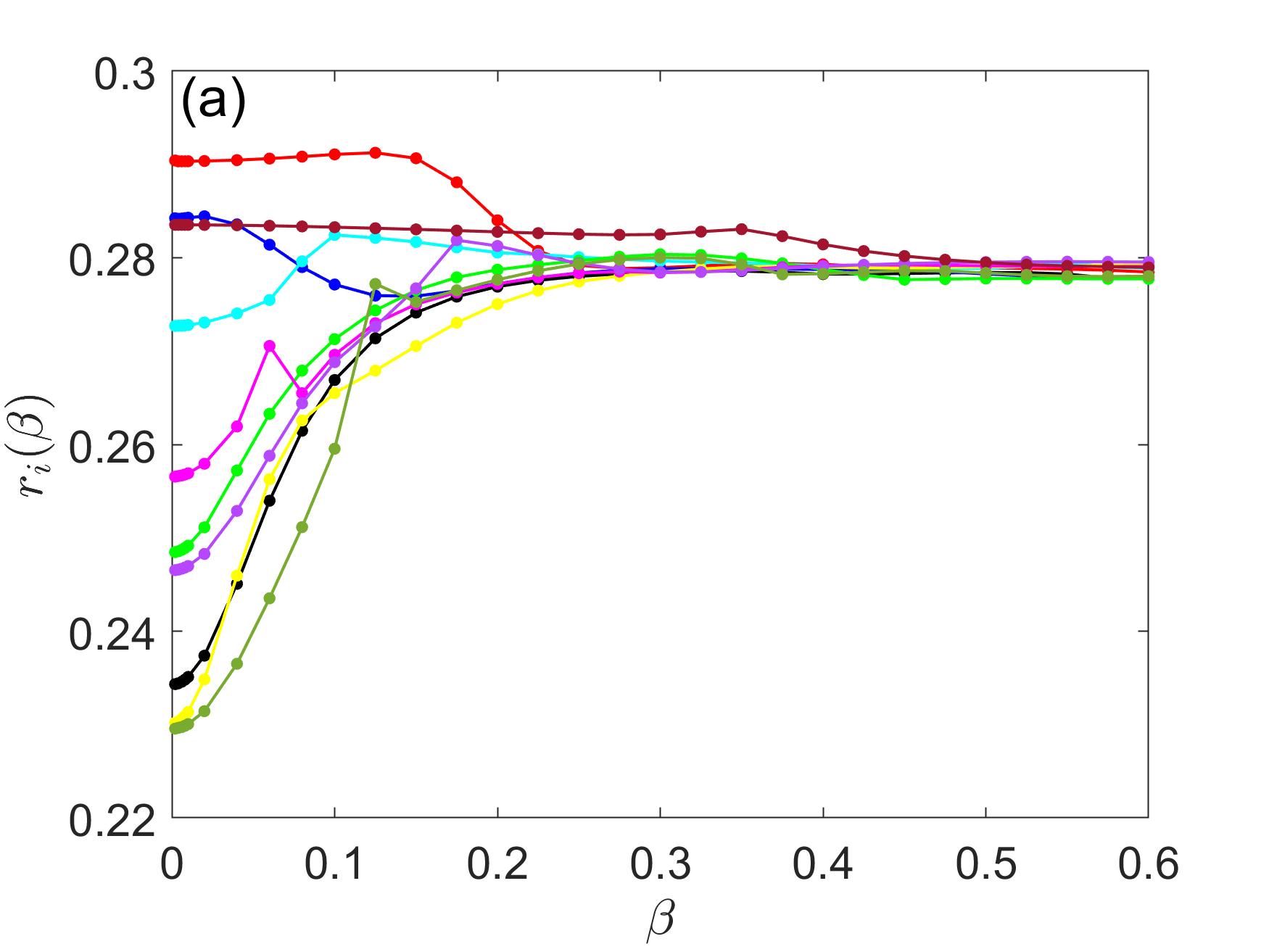}\\ 
	\includegraphics[width=\columnwidth]{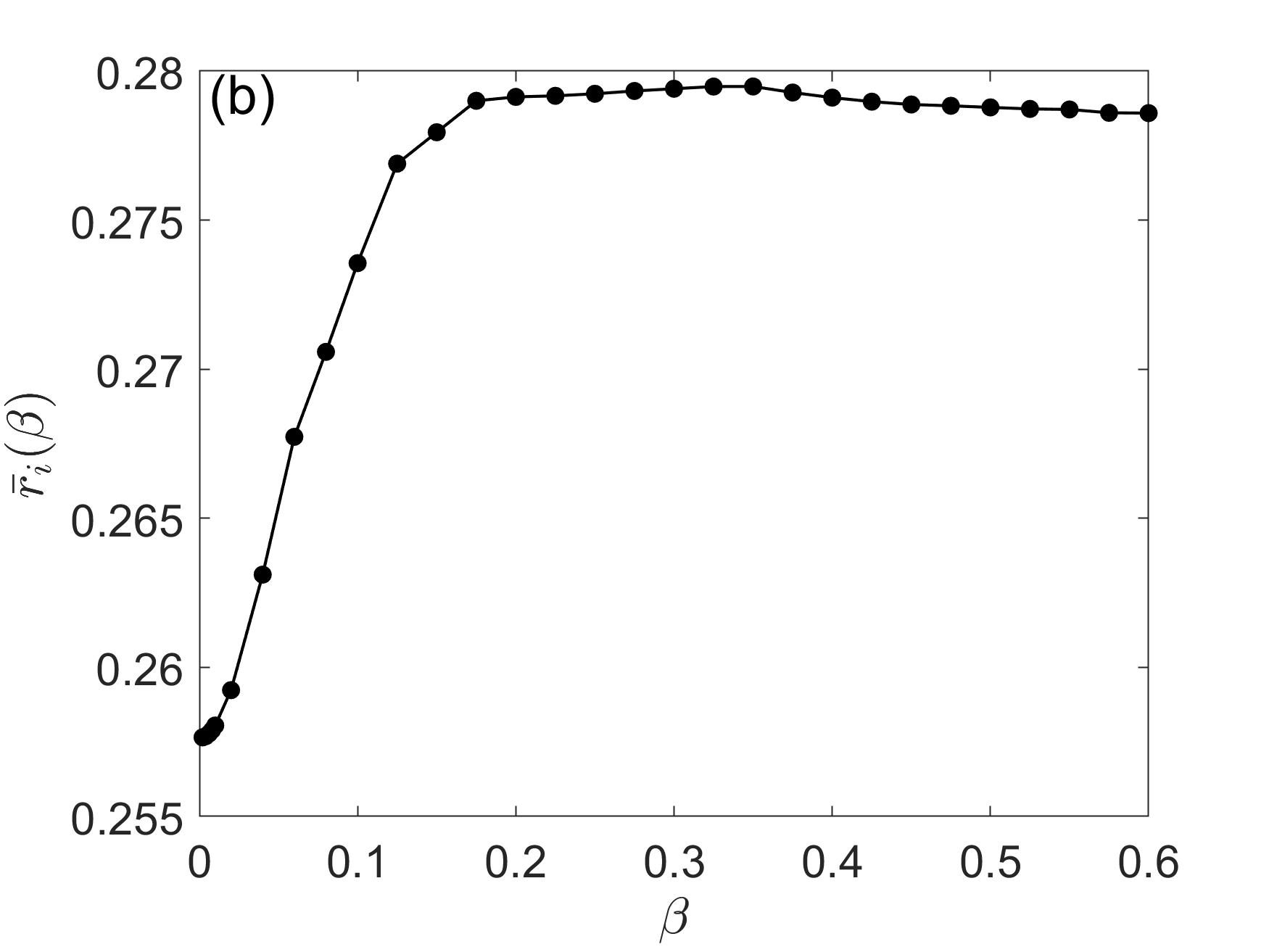}\\ 
	\caption{The surface resistance $r_i(\beta)$ at $l/\lambda=10$, $\gamma=0.04$, $n_i=1.67 \lambda^{-3}$, $\kappa=10$ and $\zeta_n=1$: (a) ten different random distribution of pins with the same density; (b) averaged $\bar{r}_i(\beta)$.  }
	\label{fig:Fig3}
\end{figure} 

The global surface resistance is calculated by solving for ${\bf u}^k(x,t)$ for each vortex moving in a particular pin configuration and averaging over all vortices according to Eqs. (\ref{p})-(\ref{rii}). Each $k$-th vortex moves in a different pinning landscape and produces a unique $r_i(\beta,k)$ which can vary significantly from vortex to vortex.  Figure \ref{fig:Fig3} shows the result of such averaging for ten different random pin distribution with the same $n_i=1.67\lambda^{-3}$. Here the low-field $r_i(\beta,k)$ fluctuate strongly but converge to the same value at high fields.  This shows that $r_i(\beta,k)$ at low fields is strongly affected by pinning, whereas $r_i(\beta,k)$ at higher fields is mostly limited by the vortex drag and the effect of pinning fluctuations weakens. The averaged $\bar{r}_i(\beta)$ shown in Fig. \ref{fig:Fig3}b first increases with $\beta$ and levels off at $\beta\gtrsim 0.2$ as the low-field $\bar{r}_i(\beta)$, which mostly results from pinning hysteretic losses, crosses over to a drag-dominated $\bar{r}_i(\beta)$. A similar low-field dependence of $R_i(H)$ has been observed on Nb cavities ~\cite{liarte,fnal_20}.

\begin{figure}[h]
	\centering
	\includegraphics[width=\columnwidth]{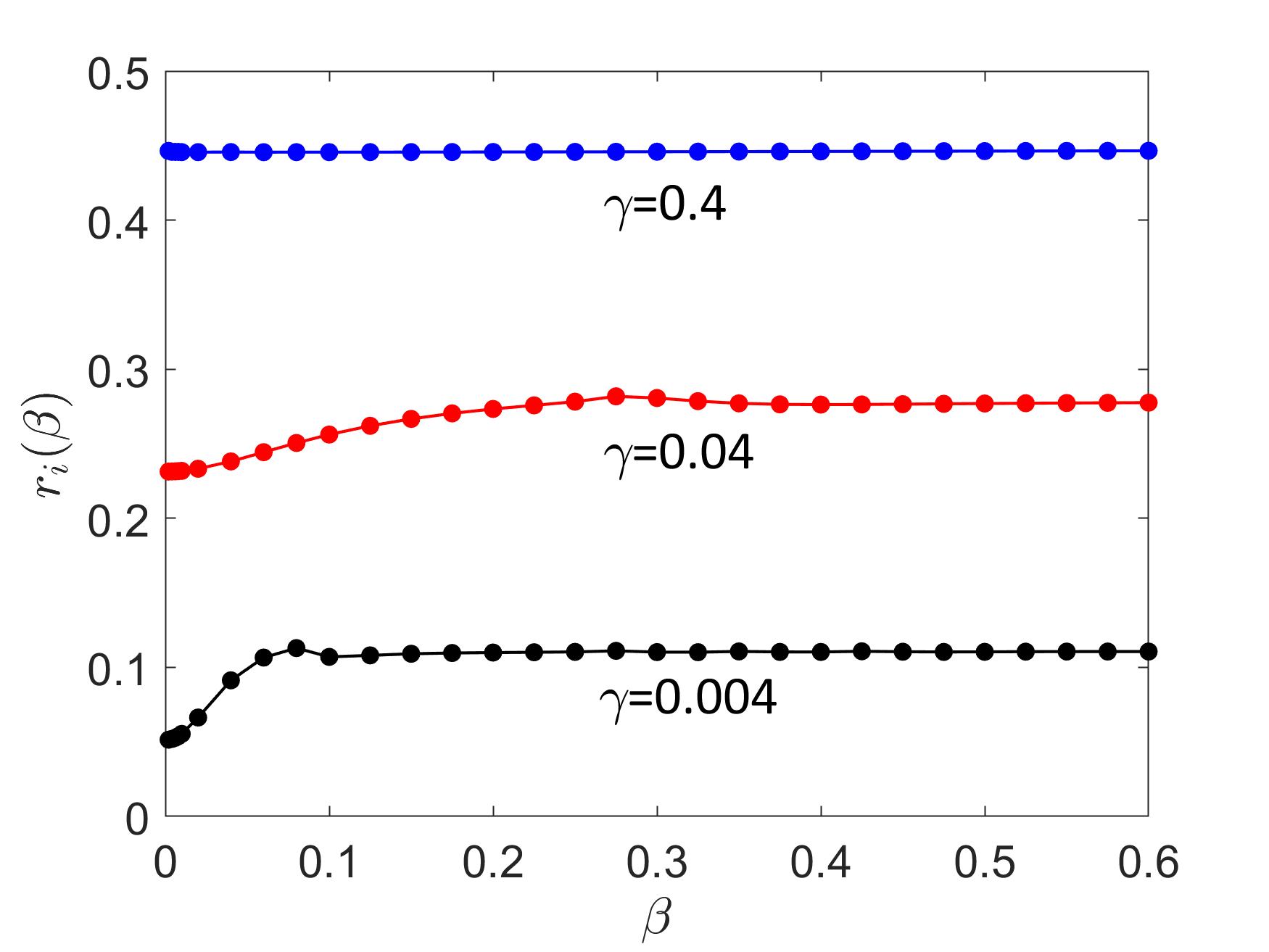}\\
	\caption{Surface resistance $r_i(\beta) $ calculated for different dimensionless frequencies $\gamma=0.004, 0.04, 0.4$. Other parameters are: $l/\lambda=10$, $n_i=0.2 \lambda^{-3}$, $\kappa=10$ and $\zeta_n=1$. }
	\label{fig:Fig4}
\end{figure}

\begin{figure}
	\centering
	\includegraphics[width=\columnwidth]{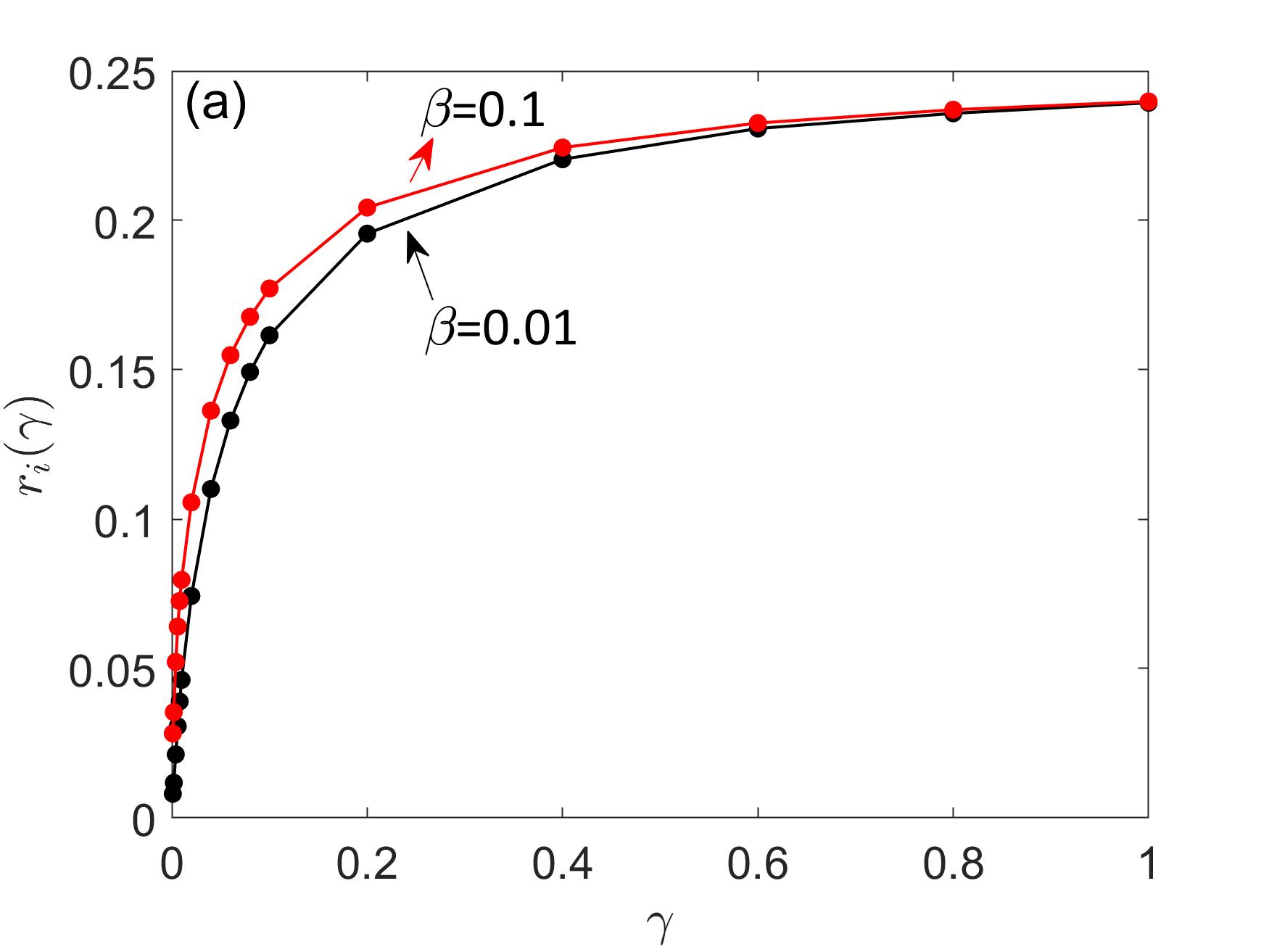}\\
	\includegraphics[width=\columnwidth]{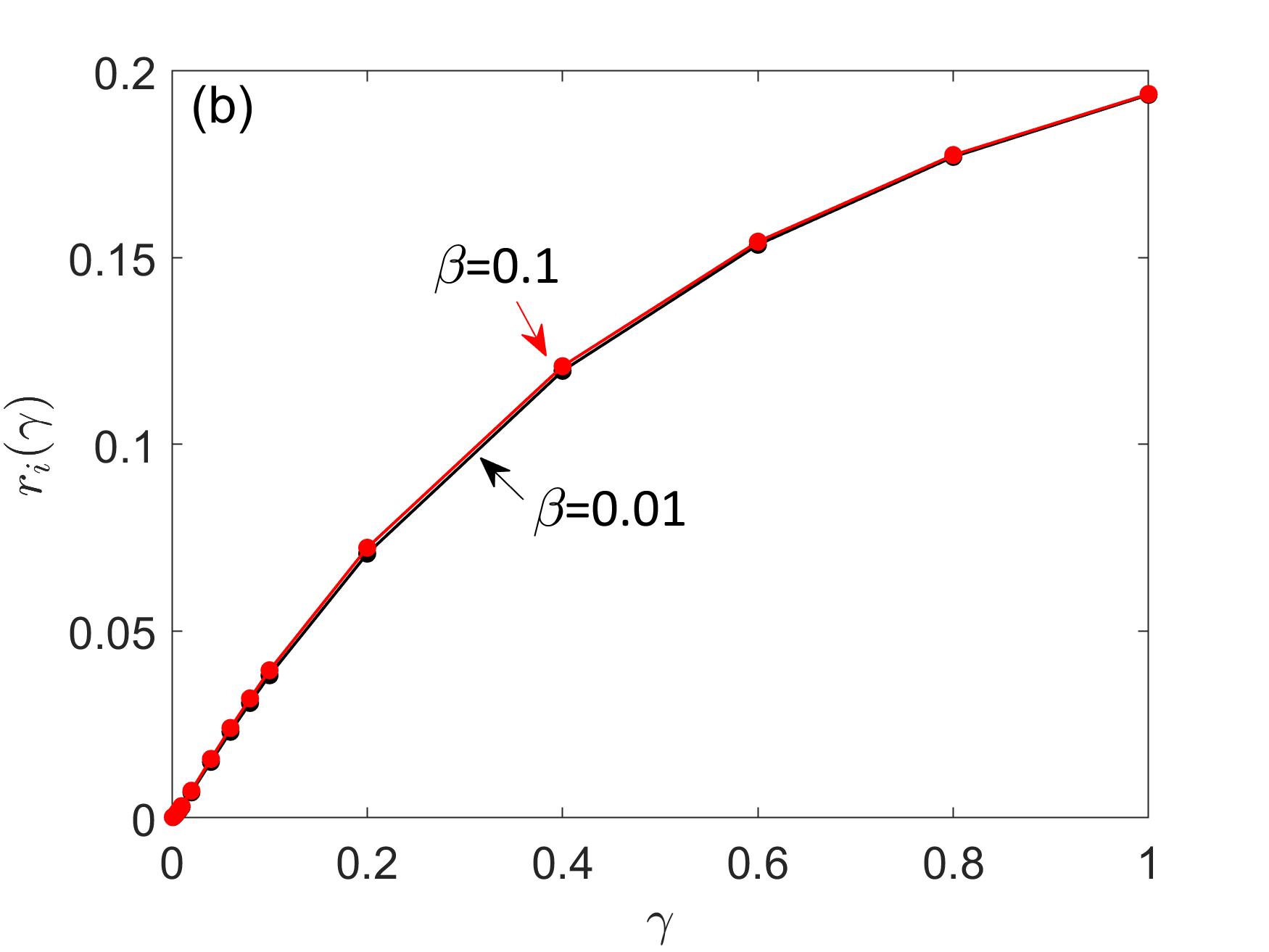}\\
	\caption{Frequency dependencies of $r_i(\gamma)$ calculated at $l/\lambda=20$, $n_i=0.0625\lambda^{-3}$, $\kappa=10$, field amplitudes 
	$\beta=0.01$ and $\beta=0.1$, and the pinning parameters (a) $\zeta_n=1$, (b) $\zeta_n=20$.}
	\label{fig:Fig5}
\end{figure}

Now we turn to the effect of pinning on the frequency dependence on $r_i(\beta,\gamma)$ shown in Fig. \ref{fig:Fig4}. At a high frequency $\gamma=0.4$ the surface resistance $r_i(\beta)$ is nearly independent of the field amplitude $\beta$ because the rf losses are dominated by the linear vortex drag. As the frequency decreases, a linear dependence of $r_i(\beta)$ develops at small fields for which pinning reduces $r_i(\beta)$. This result is consistent with the calculations of $r_i(\beta)$ in a quasi-static limit  ~\cite{liarte}. The frequency dependence $r_i(\gamma) $ is affected by both the rf field and the pinning strength. For stronger pinning ($\zeta_n=20$) represented by Fig. \ref{fig:Fig5}b, the surface resistance is affected by pinning at low frequencies and by viscous flux flow at high frequencies.  Here $r_i(\gamma)$ has a linear frequency dependence at small $\gamma$, which is indicative of hysteretic losses of an elastic vortex driven through a strong random pinning potential ~\cite{ce,cam3,cam4,liarte}, unlike  $r_i\propto\gamma^2$ characteristic of reversible small vibrations of a vortex around the equilibrium position ~\cite{tf1,ag_sust}. The linear frequency dependence of $r_i(\gamma)$ is not described by the Gittleman-Rosenbluth model ~\cite{Gittleman} in which a vortex in a thin film is modeled by an overdamped particle in a parabolic pinning potential.

The frequency dependence of $r_i(\gamma)$ changes as pinning becomes weaker. Shown in Fig. \ref{fig:Fig5} is $r_i(\gamma)$ calculated for two values of the pinning parameter, $\zeta_n=1$ and $\zeta_n=20$. At high frequencies $r_i(\gamma)$ tends to the same value $\simeq 0.25$ dominated by the vortex drag for both $\zeta_n=20$ and $\zeta_n=1$. However, $r_i(\gamma)$ for $\zeta_n=1$ decreases sharply at small $\gamma$, rather different from the linear $r_i(\gamma)$ at $\zeta_n=20$. Moreover, $r_i(\gamma)$ for $\zeta_n=1$ does not vanish at $\gamma\to 0$ if the field amplitude $H$ exceeds a critical value $H_p$. At $H>H_p$ the Lorentz force $H\phi_0$ exceeds the maximum pinning force of the vortex segment of length $l$ and the vortex starts moving along the film under a parallel magnetic field exceeding the depinning field $H_p(l,\omega)$.

\subsection{RF depinning field}

\begin{figure}[h]
	\centering
	\includegraphics[width=\columnwidth]{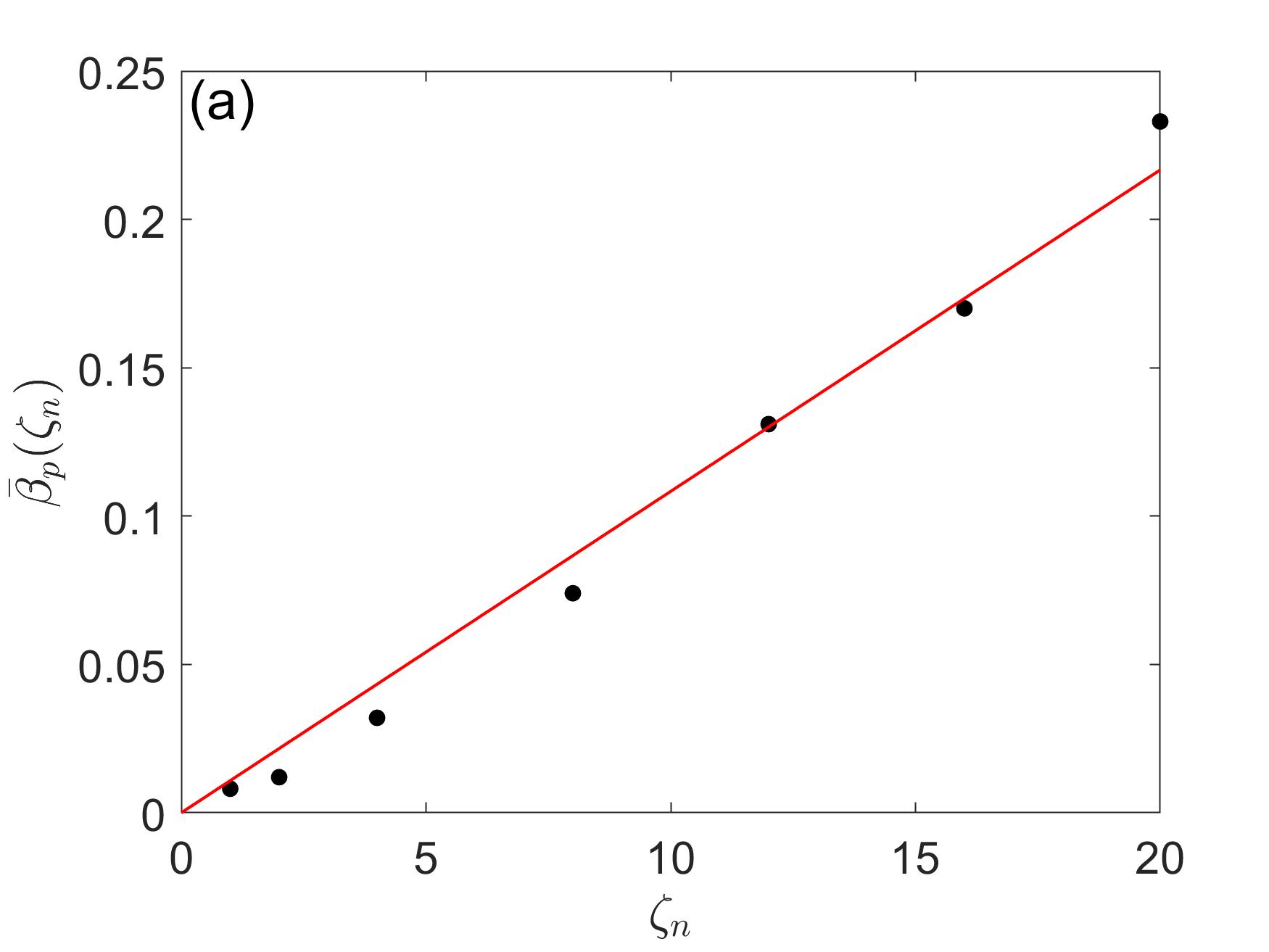}\\
	\includegraphics[width=\columnwidth]{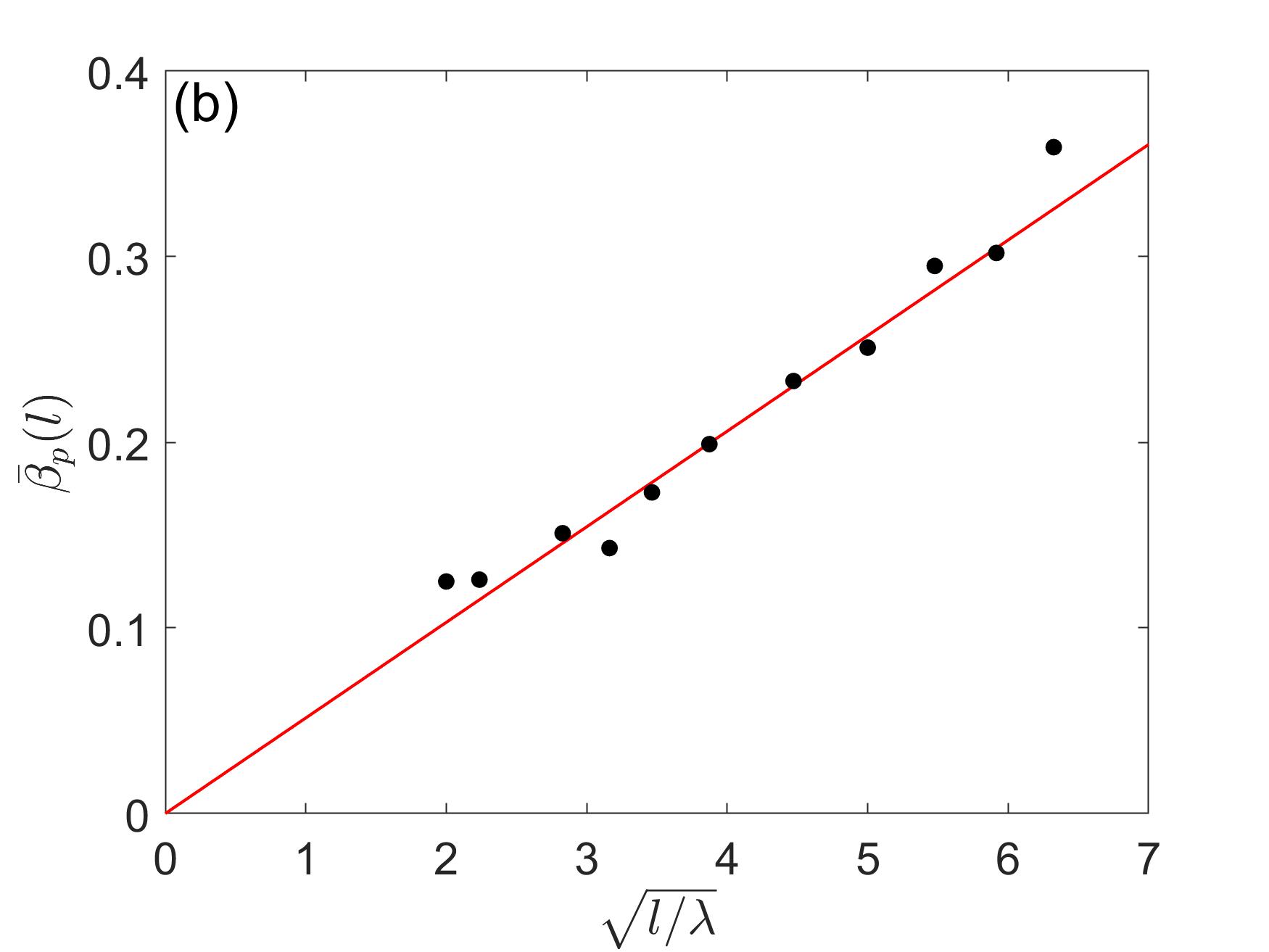}
	\caption{ Depinning field $\beta_p$ as a function of: (a) the pinning parameter $\zeta_n$ at $l/\lambda=20$; (b) the vortex length $l$ at 
	$\zeta_n=20$.  Here $ \beta_p$ was calculated 
	by averaging over 10 random pin distributions with $n_i=0.0625\lambda^{-3}$ at $\gamma=0.001$ and $\kappa=10$. The red lines show $\beta_p$ 	given by Eq. (\ref{bc}).}
	\label{fig:Fig6}
	\end{figure}

The depinning field $\beta_p(l)=H_p/H_{c1}$ of a single vortex can 
be evaluated using the collective pinning theory \cite{blatter,ehb}. If the film thickness $l$ is small enough (see below), 
pinning mostly causes displacements of a rigid vortex which remains nearly straight and
perpendicular to the film surface. Such a vortex interacts with
$N_p\simeq r_p^2l/l_i^3 \gg 1$ pins, adjusting its position in such a way that the net
pinning force vanishes, where $r_p\sim\xi$ is the pin interaction radius and $l_i$ is the pin spacing $(r_p,l_i)\ll l$. 
The vortex gets depinned if the Lorentz force $\phi_0H$ exceeds the net pinning force
$f_p\sqrt{N_p}$ from uncorrelated pins in the volume $r_p^2l$. The condition $\phi_0H_p\simeq f_p\sqrt{N_p}$ yields
    \begin{equation}
    H_p\simeq \frac{u_p}{\phi_0}\sqrt{n_i l}. 
    \label{Ic}
    \end{equation}
Here $n_i = l_i^{-3}$ is a volume pin density, and $u_p=f_pr_p=\pi\xi U_n$ is the elementary
pinning energy for $U(x,{\bf u})$ given by Eq. (\ref{eq3}). The dimensionless depinning field $\beta_p=H_p\phi_0/\epsilon$ expressed in terms of the pinning parameter $\zeta_n=2\kappa^2 u_p/\pi\epsilon\xi$ is given by:
\begin{equation}
\beta_p\simeq (\zeta_n\lambda/\kappa^2)\sqrt{n_il}.
\label{bc}
\end{equation}
This $\beta_p(\zeta_n,l)$ of the collective pinning theory is in good agreement with the calculated $\beta_p(\zeta_n,l)$ 
shown in Fig. \ref{fig:Fig6}. 
  
As the film thickness increases, a gradual transition from the 2d pinning of a rigid vortex to the 3d
pinning of a deformable vortex is expected to occur if $l$ exceeds the Larkin pinning length $L_c$. 
Here $L_c$ can be evaluated using the collective pinning theory ~\cite{blatter}:
    \begin{equation}
    H_p\sim\frac{u_p}{\phi_0}\sqrt{n_iL_c},
    \qquad \frac{\epsilon\xi}{L_c} \sim \phi_0H_p.
    \label{3d}
    \end{equation}
The first relation is similar to Eq. (\ref{Ic}) for a vortex segment
of length $L_c$, and the second one reflects the balance of the
total Lorentz and bending elastic forces acting on the vortex segment
displaced by $\simeq \xi$ from its equilibrium position. 
From Eq. (\ref{3d}), it follows that $L_c\simeq (\epsilon\xi/u_p)^{2/3}n_i^{-1/3}$ and:
\begin{equation}
L_c\sim (2\kappa^2/\zeta_n)^{2/3}n_i^{-1/3}.
\label{lc} 
\end{equation}
For $\kappa=10$, $\zeta_n=20$ and $n_i=0.0625\lambda^{-3}$,
we get $L_c \sim 5\lambda$. This estimate of $l>L_c$ above which $\beta_p(l)$ is expected to 
level off is inconsistent with the numerical results shown in Fig. \ref{fig:Fig6}b, where 
the square root dependence $\beta_p\propto \sqrt{l}$ persists all the way to $l\simeq 50\lambda$, even though  
$\zeta_n=2\kappa^2 u_p/\pi\epsilon\xi=20$ at $\kappa=10$ corresponds to strong core pinning with 
$u_p\simeq 0.3\epsilon\xi$. Such pin in a sparse flux line lattice (FLL) would cause strong bending distortions of the vortex and its hysteretic pinning response since the Labusch criterion of weak pinning is not satisfied ~\cite{sp1,sp2,sp3,sp4,sp5}. However, as shown in Appendix B, the distinction between strong (hysteretic) and weak (non hysteretic) pinning based on the Labusch criterion for a bulk  FLL \cite{sp1,sp2,sp3,sp4,sp5} is not applicable to a single vortex in a film. 

The 2d collective pinning theory in a film with $l>L_c$ remains applicable because the vortex can adjust its position to reduce bending distortions. As result, pinning response of the vortex interacting with a single pin in a film is non-hysteretic, no matter how strong $f_p$ is (see Appendix B). This behavior is different from the hysteretic pinning response of a vortex in an infinite FLL, where the ends of a long vortex are fixed by its equilibrium position in the FLL ~\cite{sp1,sp2,sp3,sp4,sp5}. Pinning response of the vortex in a film may become hysteretic only if several pins are involved. Because the vortex can adjust is positions in a film and remain nearly straight, the 2d collective pinning may persist even at $l\simeq 10L_c$, as our results show. The transition to the 3d pinning will eventually occur at larger $l$ which is outside the limited range of $l<50\lambda$ in our simulations. In addition, the transition from 2d to 3d collective pinning is a gradual change in $J_c(l)$ which can also depend on a correlation function of pin distribution \cite{brandt1,ag_pin}. 

\subsection{Surface pinning} \label{surface} 

In the case of surface pinning caused by segregation of materials defects at the surface ~\cite{clare} the Lorentz and pinning forces are only applied to the tip of the vortex the rest of which can move freely (see Fig. 1b). The surface pinning is modeled here by a random distribution of pins in a layer of thickness $\lambda$.
\begin{figure}[ht]
	\centering
	\includegraphics[width=\columnwidth]{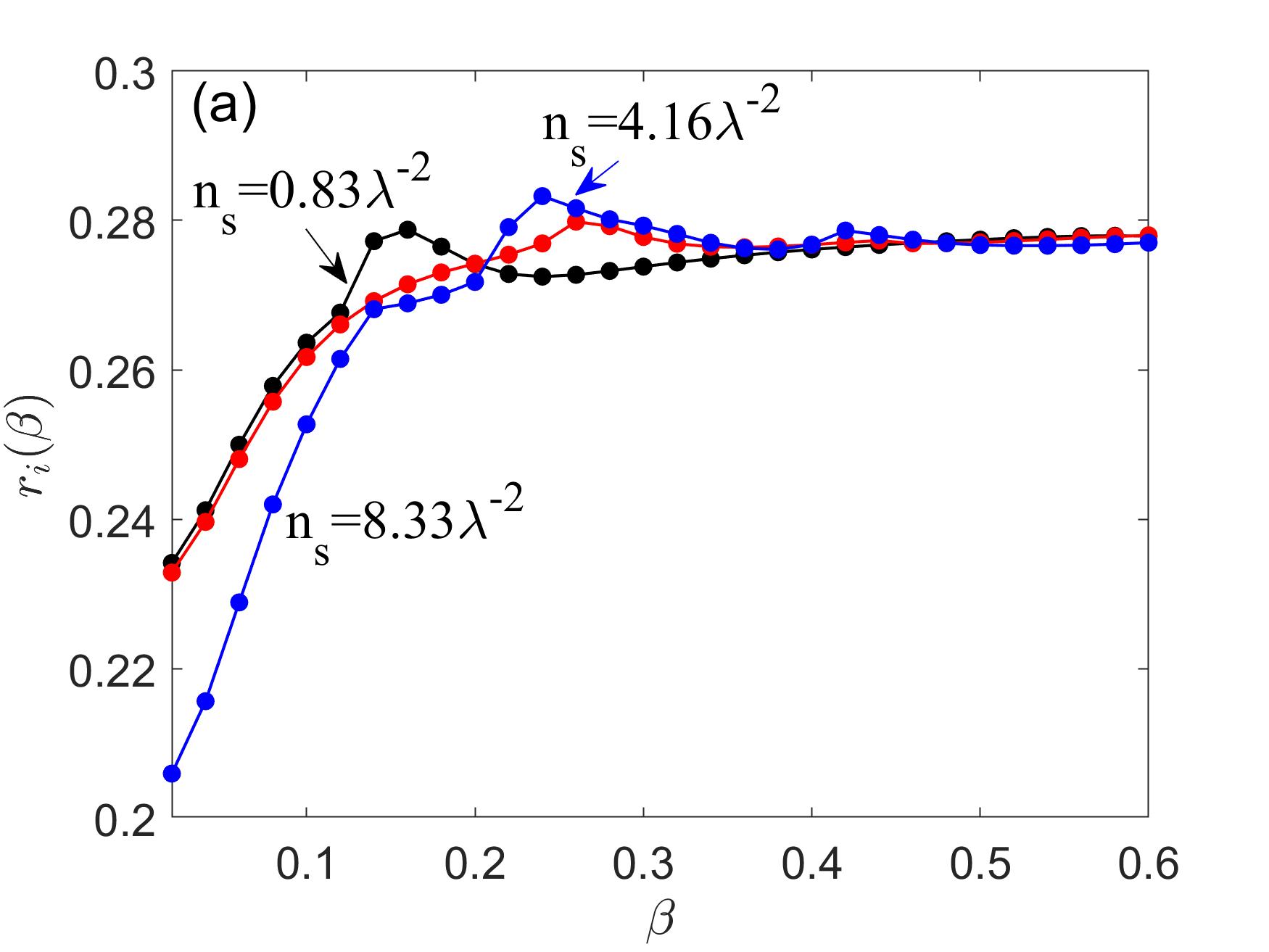}\\
	\includegraphics[width=\columnwidth]{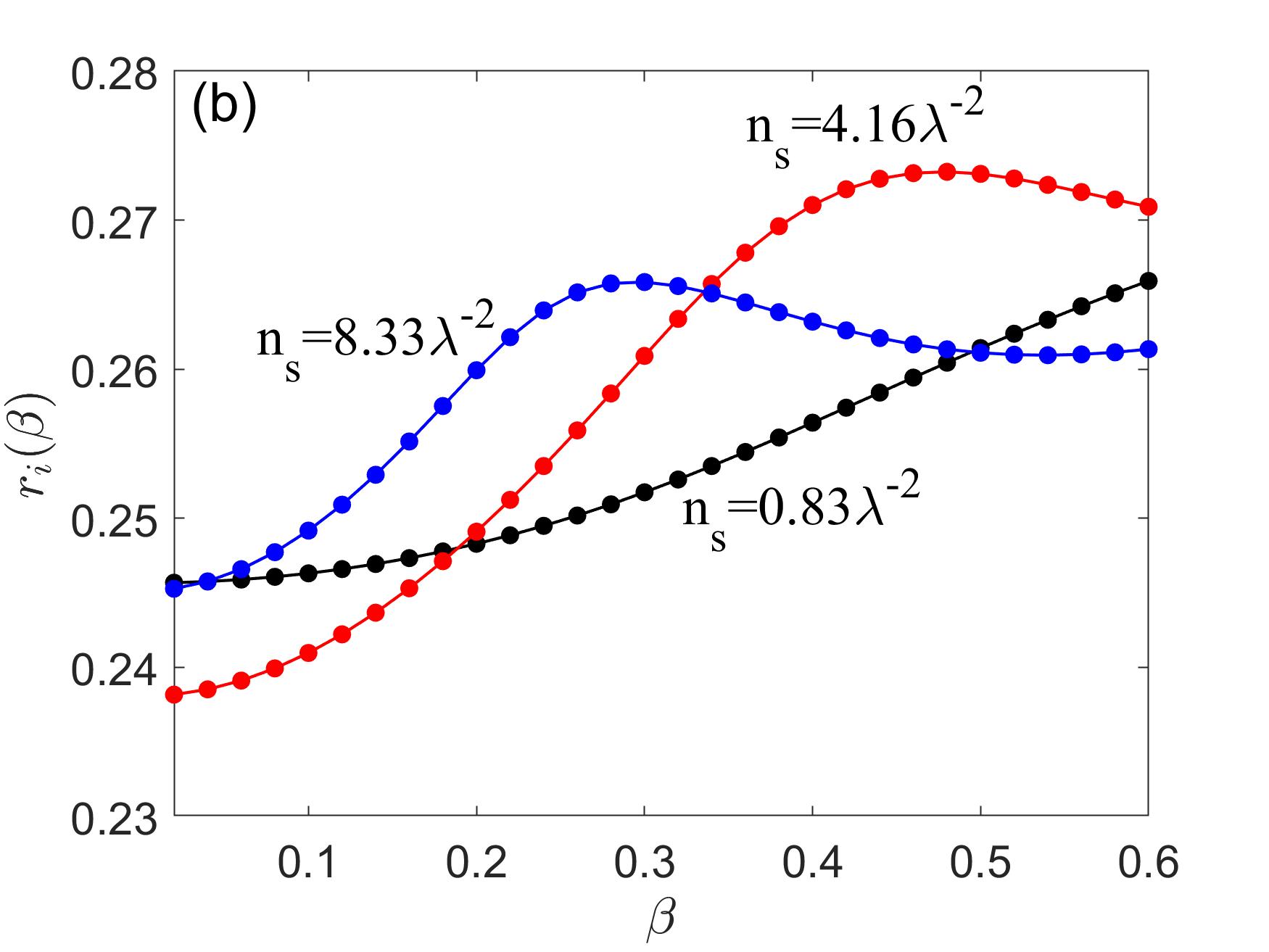}\\ 
	\caption{$r_i(\beta)$ for surface pinning calculated at $l/\lambda=10$, $\gamma=0.04$, $n_s=0.83\lambda^{-2}, 4.16\lambda^{-2}, 8.33\lambda^{-2}$: (a) $\kappa=10$ and $\zeta_n=1$, (b) $\kappa=2$ and $\zeta_n=0.08$.}
	\label{fig:Fig7}
	\end{figure} 	

Shown in Fig. \ref{fig:Fig7} are $r_i(\beta) $ calculated for different values of the GL and pinning parameters $\kappa$ and $\zeta_n$, and sheet pin densities $n_s=0.83\lambda^{-2}, 4.16\lambda^{-2}, 8.33\lambda^{-2}$. 
At $\kappa=10$ and $\zeta_n=1$ the surface resistance $r_i(\beta)$ fluctuates strongly, depending on a pin configuration.  At $\kappa=2$ the pinning parameter $\zeta_n$ is smaller and the effect of fluctuations diminishes because the elementary pinning potentials overlap.  As $\beta$ increases the amplitude of vortex swings increases, so the vortex probes different regions of the pinning potential.

\begin{figure}[ht]
	\centering
	\includegraphics[width=\columnwidth]{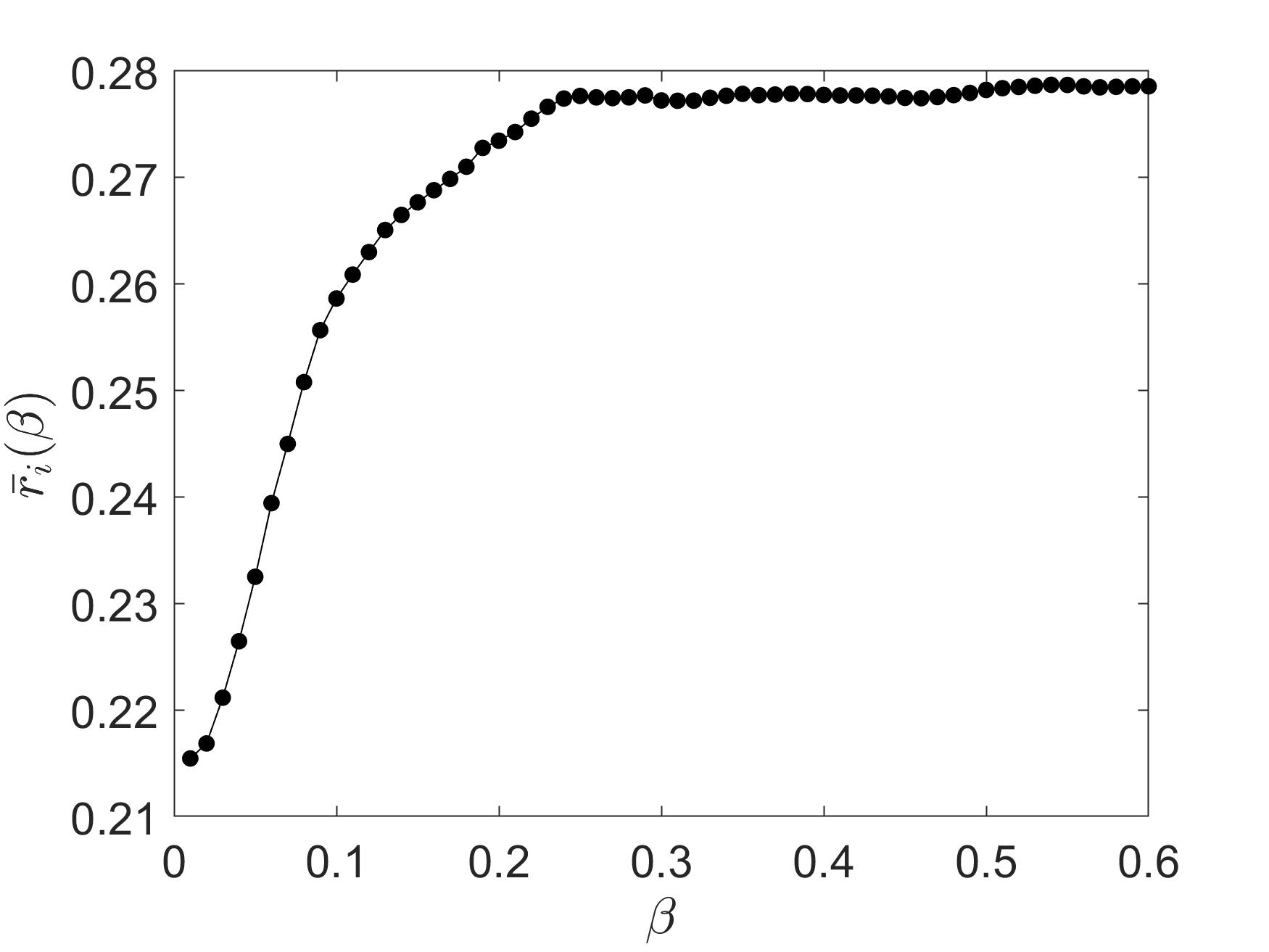}\\ 
	\caption{$\bar{r}_i(\beta)$ for surface pinning averaged over ten different random distribution with $n_s=8.33\lambda^{-2}$. Here $\bar{r}_i(\beta)$ 		was calculated at $l/\lambda=10$, $\gamma=0.04$, $\kappa=10$, and $\zeta_n=1$. }
	\label{fig:Fig8}
\end{figure}

The mean $\bar{r}_i(\beta)$ obtained by averaging $r_i(\beta)$ over ten different pin configurations with the same sheet density is shown Fig. \ref{fig:Fig8}. The field dependence of $\bar{r}_i(\beta)$ for surface pinning is similar to $\bar{r}_i(\beta)$ for bulk pinning shown in Fig. \ref{fig:Fig3}b.  In both cases the low-field behavior of $\bar{r}_i(\beta)$ is affected by pinning. As $\beta$ increases, $\bar{r}_i(\beta)$ levels off at a constant value dominated by the vortex drag.

\begin{figure}[ht]
	\centering
	\includegraphics[width=\columnwidth]{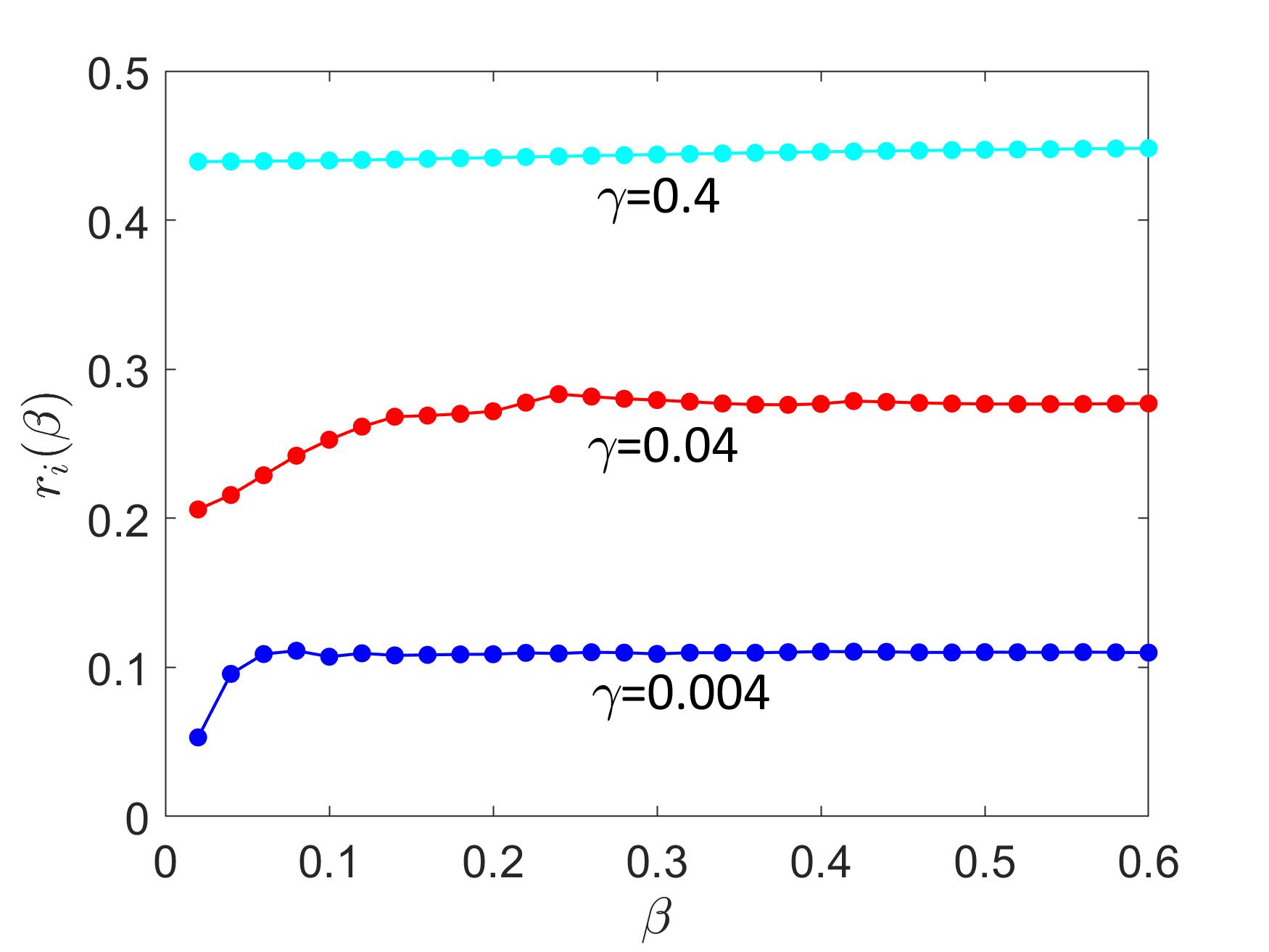}\\
	\caption{$r_i(\beta)$ for surface pinning calculated at different frequencies: (a) $\gamma=0.004, 0.04, 0.4$, $l/\lambda=10$, $n_s=2.5 \lambda^{-2}$, 
	$\kappa=10$ and $\zeta_n=1$.}
	\label{fig:Fig9}
\end{figure}

The effect of frequency $\gamma=f/f_0$ on the field-dependence of $r_i(\beta)$ is shown in Fig. \ref{fig:Fig9}. At $\gamma=0.4$ the surface resistance is dominated by the vortex drag and is nearly independent of $\beta$. As $\gamma$ decreases a dip in $r_i(\beta)$ develops at low fields, where the surface resistance is affected by pinning. At the lowest frequency $\gamma=0.004$ the elastic skin length $L_\omega\simeq \lambda/\sqrt{2\pi\gamma}\simeq 6.3\lambda$ is comparable to the vortex length $l=10\lambda$ so the vortex oscillates as a nearly rigid rod. The overall effect of frequency on $r_i(\beta)$ for surface pinning is similar to that of $r_i(\beta)$ for bulk pinning shown in Fig. \ref{fig:Fig4}.

\subsection{Cluster pinning} \label{cluster}

Pinning by clusters of materials defects such as impurities of nanoprecipitates ~\cite{clare} depicted in  Fig. \ref{fig:Fig1}c has the following distinctive features as compared to the bulk and surface pinning: 1. At $H=0$, a tilted trapped vortex can be pinned by misaligned clusters; 2. The vortex tip at the surface exposed to the rf field can get de-pinned as $H$ increases and the vortex straightened out.   
We consider pins distributed randomly in two $\lambda\times0.5\lambda\times0.5\lambda$ clusters on both surfaces of the film of thickness $l=10\lambda$. The clusters with $n_i=60\lambda^{-3}$ each were shifted with respect to each other by $\lambda/2$ along $y$ and $z$ as shown in Fig. \ref{fig:Fig1}c.
 
\begin{figure}[ht]
	\centering
	\includegraphics[width=\columnwidth]{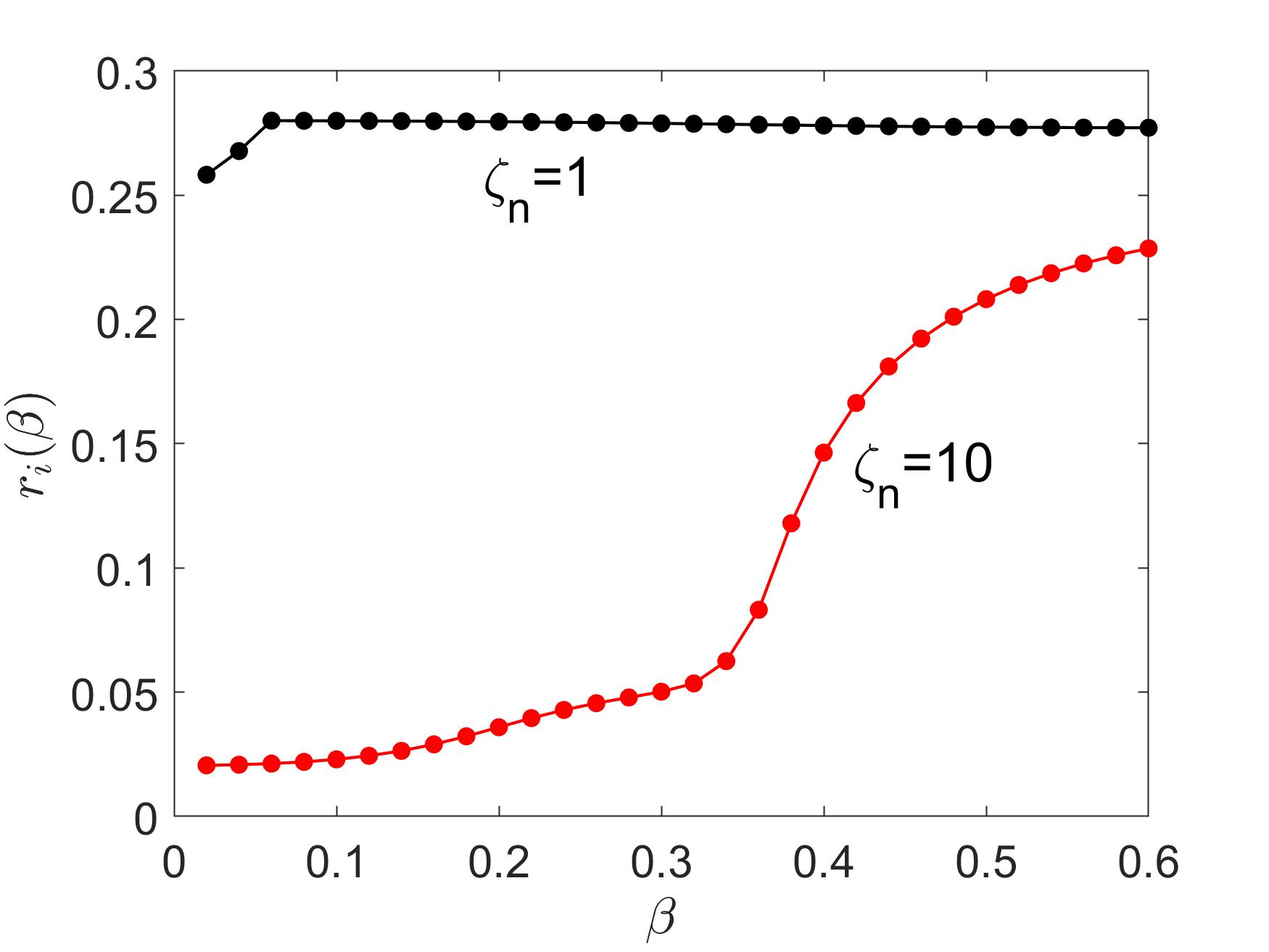}\\ 
	\caption{$ r_i(\beta) $ for cluster pinning calculated at two pinning parameters $\zeta_n=1$ and $\zeta_n=10$, $l/\lambda=10$, $\gamma=0.04$, 		and $\kappa=10$.}
	\label{fig:Fig10}
\end{figure} 

\begin{figure}[ht]
	\centering
	\includegraphics[width=\columnwidth]{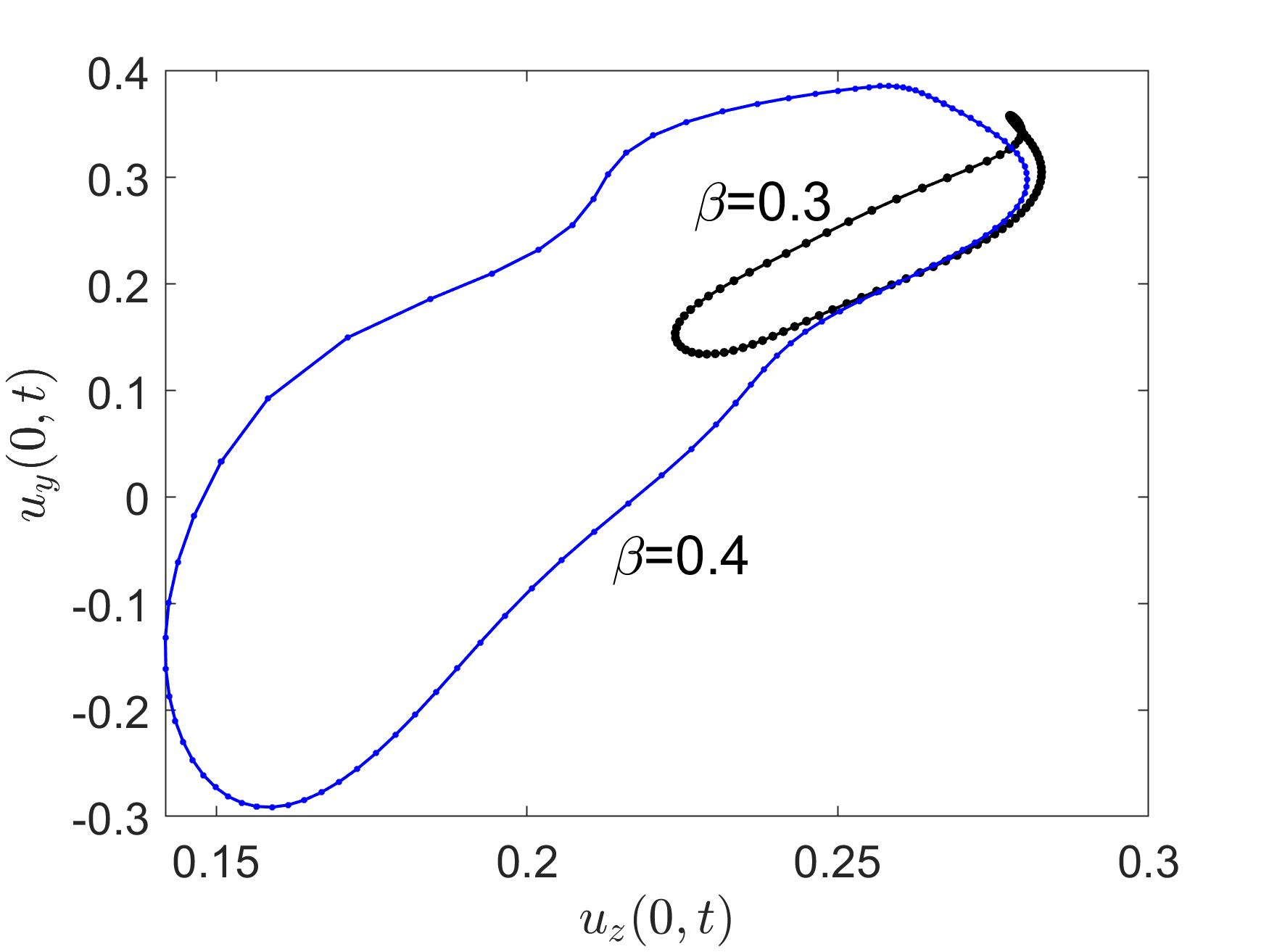}\\
	\caption{Trajectories of the vortex tip at $\beta=0.3,0.4$ for $l/\lambda=10$, $\gamma=0.04$, $\kappa=10$ and $\zeta_n=10$. }
	\label{fig:Fig11}
\end{figure}

Figure \ref{fig:Fig10} shows the field dependencies of $r_i(\beta)$ for cluster pinning calculated at two values of the pinning parameter $\zeta_n$. The surface resistance $r_i(\beta)$ for stronger pinning with $\zeta_n=10$, increases sharply above $\beta\simeq 0.35$ due to the temporal escape of the vortex tip from the pin cluster, while the other end of the vortex remains pinned. The temporal depinning of the vortex tip occurs during a portion of the rf period at which $H(t)=H\sin\omega t$ exceeds the depinning field $H_p$, resulting in large-amplitude swings of the vortex and a sharp increase in $r_i(\beta)$. For weaker pinning $(\zeta_n=1)$, the vortex tip escapes the cluster at a lower field $\beta\gtrsim 0.04$, consistent with $\beta_p\propto \zeta_n$ shown in Fig. \ref{fig:Fig6}. The change in the trajectory of the vortex tip as $\beta$ exceeds $\beta_p\simeq 0.35$ is shown in Fig. \ref{fig:Fig11}. Temporal depinning and straightening of the vortex at $\beta(t)>\beta_p$ can produce a hysteretic field dependence of $r_i(\beta)$ upon increase and decrease of the field amplitude.

\begin{figure}[]
	\centering
	\includegraphics[width=\columnwidth]{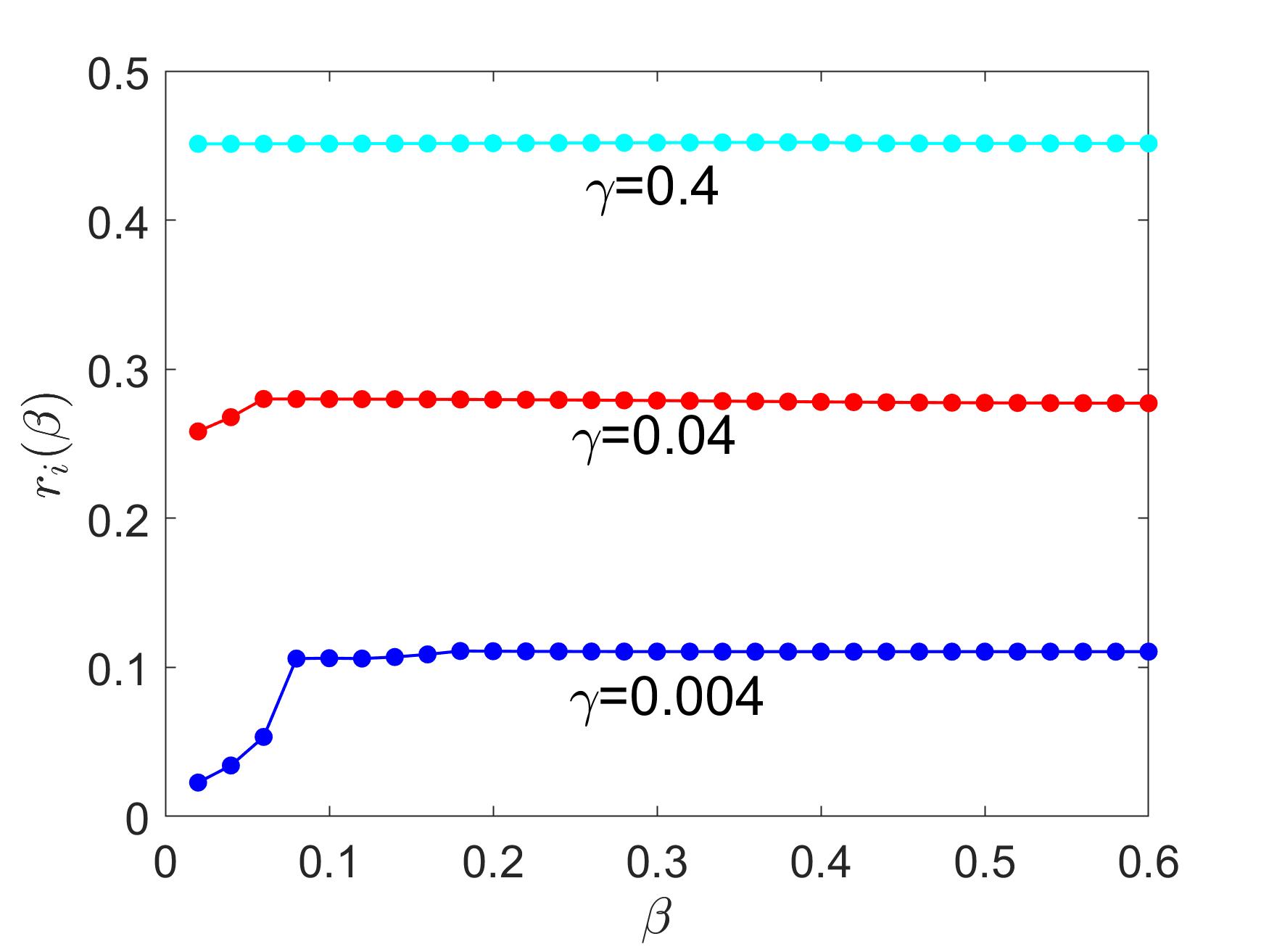}\\
	\caption{$r_i(\beta) $ vs $\beta$ calculated at  $\gamma= 0.004, 0.04, 0.4$, $l/\lambda=10$ for the case of $\kappa=10$ and $\zeta_n=1$.}
	\label{fig:Fig13}
\end{figure} 

Shown in Fig. \ref{fig:Fig13} is the effect of frequency on $r_i(\beta)$. At high frequencies $(\gamma>0.4)$, the surface resistance is controlled by the vortex drag and is practically independent of $\beta$. The low-field dip in $r_i(\beta)$ caused by pinning becomes more pronounced at lower frequency $\gamma=0.004$ for which a cusp at $\beta\approx 0.08$ results from the temporal escape of the vortex tip from the pin cluster. Except for this cusp feature, the effect of frequency on $r_i(\beta)$ is qualitatively similar to the results shown in Figs.  \ref{fig:Fig4} and \ref{fig:Fig9} for bulk and surface pinning.

\section{Larkin-Ovchinnikov Instability} \label{Larkin} 

At strong rf fields the decrease of $\eta(v)$ with $v$ should be taken into account. Addressing this effect for a vortex driven by a surface Meissner current raises the following issues: 
1. What happens if a tip of the vortex moves faster than $v_0$ while the rest of the vortex does not?  2. How is the LO instability affected by pinning for the geometry shown in Fig 1? 3. How could the decrease of $\eta(v)$ with $v$ manifest itself in the dependencies of $R_i(H,\omega)$ on $H$, $\omega$ and pinning strength? For a vortex segment pinned by a single strong pin in the bulk, some of these issues were addressed previously ~\cite{Manula}, and the effect of artificial pinning centers on the LO instability in films under a dc magnetic field and transport current was investigated in Refs. \onlinecite{lop1,lop2}. Here we calculate $R_i(H,\omega)$, taking into account the LO velocity dependence of $\eta(v)$ and solving Eqs. (\ref{dyneq1})-(\ref{bc0}) by Matlab ~\cite{Matlab} and the method of lines ~\cite{Methodofline}. Below we show a few representative examples of $R_i(H,\omega)$, relegating to Ref. \onlinecite{SI} for more details. 

Given the lack of experimental data on $v_0(T)$ at low temperatures, we solved Eqs. (\ref{dyneq1})-(\ref{bc0}) for different frequencies $\gamma$ and the LO parameters $\alpha=\alpha_0\gamma^2=1$ and $\alpha=10$. For these values of $\alpha$, the critical velocity $v_0=f_0\lambda/\sqrt{\alpha_0}$ may cover $v_0\sim 1$ km/s near $T_c$ and take into account the observed decrease of $v_0(T)$ at low $T$ ~\cite{lo3,lo9} (see also Eqs. (\ref{vol})-(\ref{taup})).  Here $R_i$ was calculated for bulk pinning at $\zeta_n=0.08$ and $\zeta_n=0.8$ at $\kappa=2$. 

\begin{figure}[h!]
	\centering
	\includegraphics[width=\columnwidth]{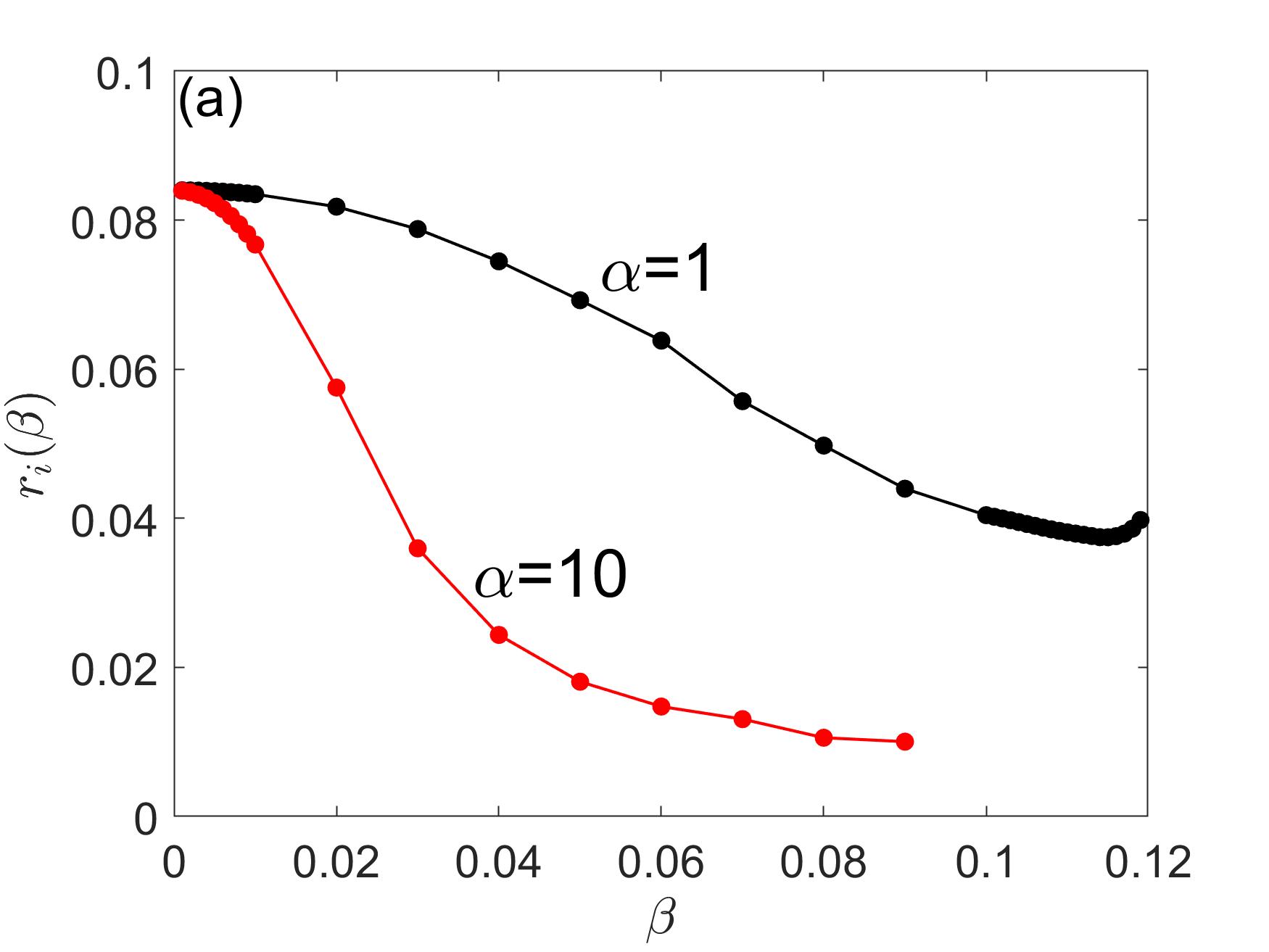}
	\includegraphics[width=\columnwidth]{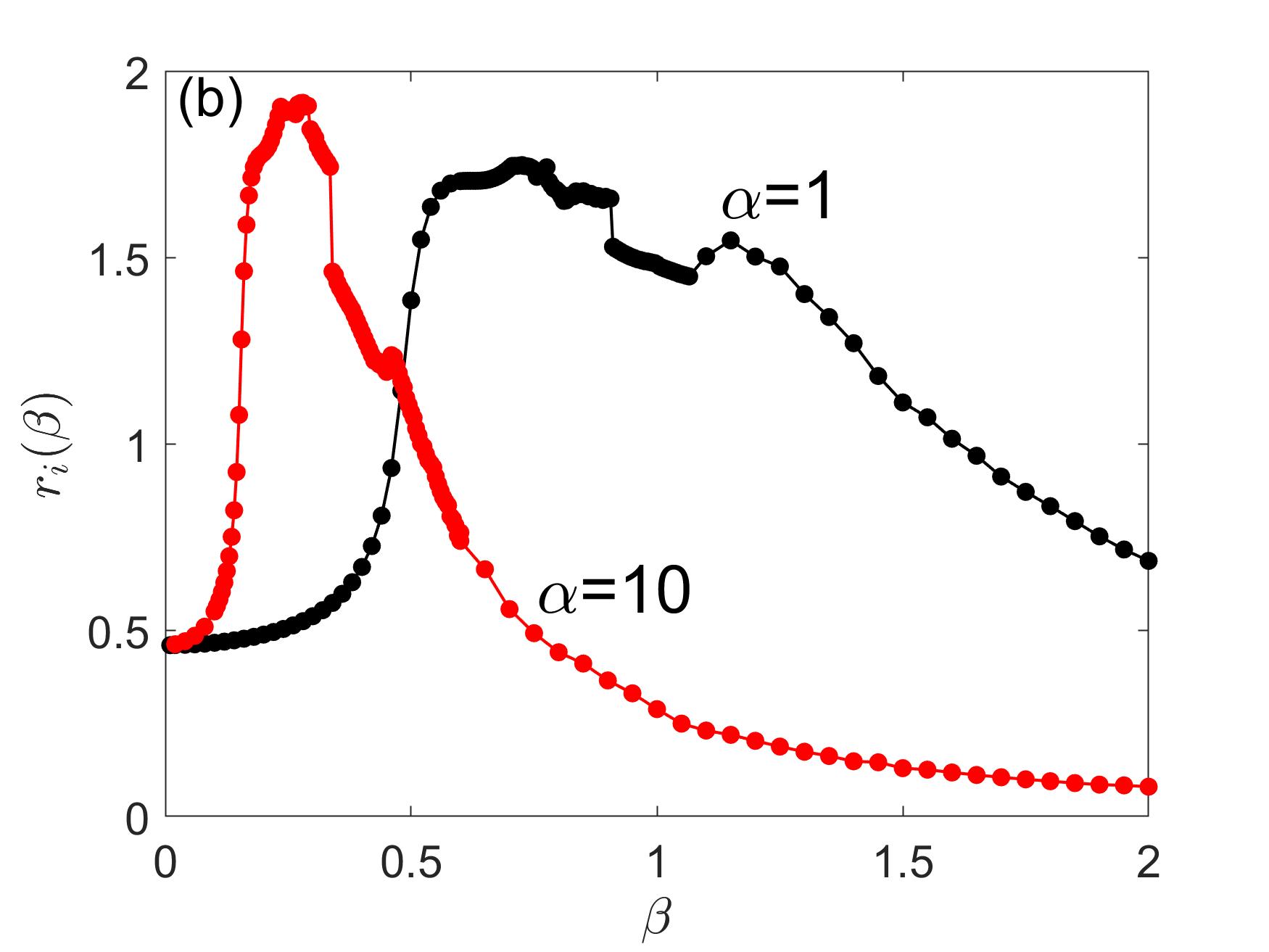}
	\includegraphics[width=\columnwidth]{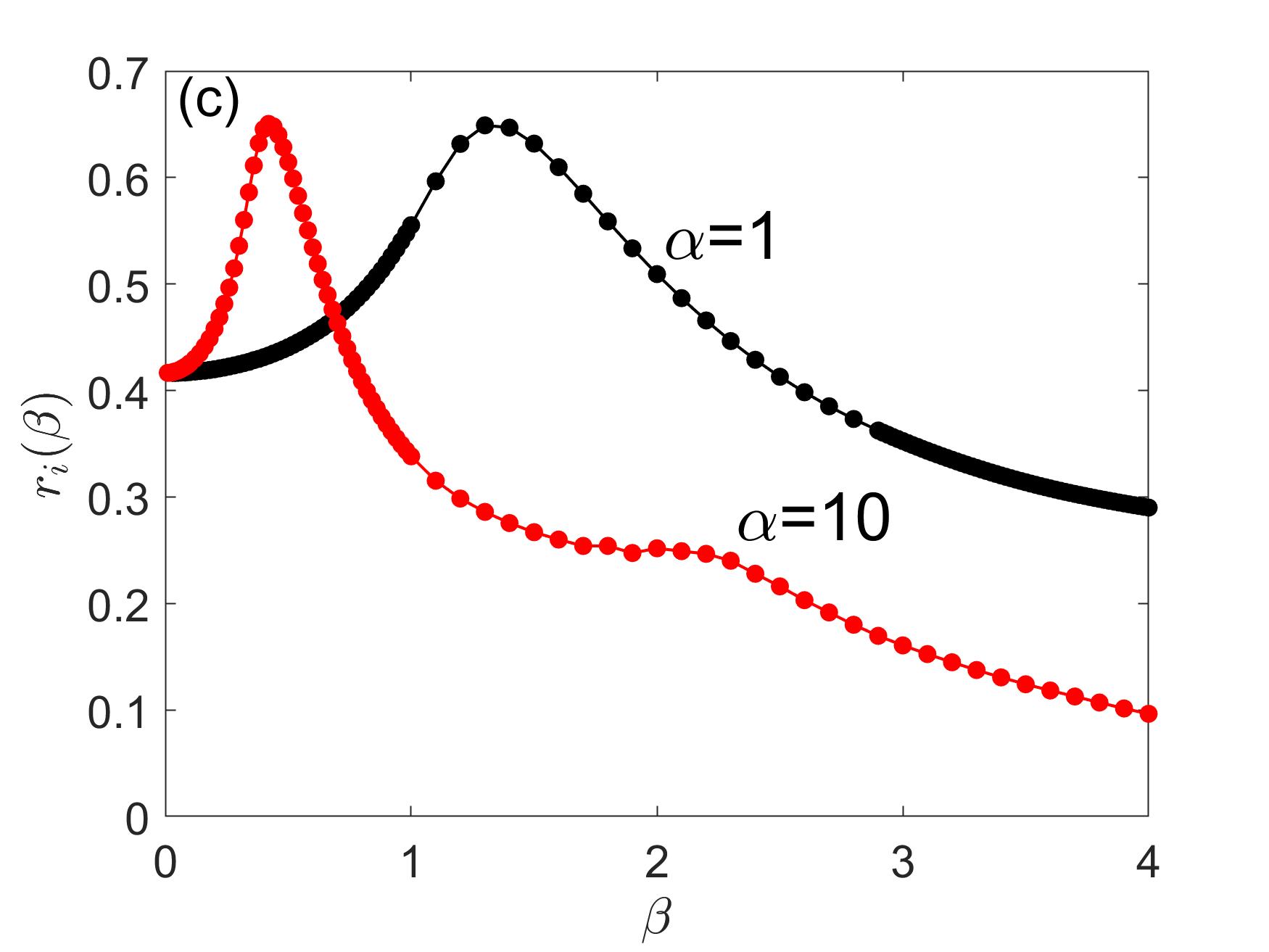}
	\caption{$r_i(\beta) $ calculated at $l/\lambda=10$, $\kappa=2$, for bulk pinning with $\zeta_n=0.8$, $n_i=0.25 \lambda^{-3}$ and: (a) $\gamma=0.01$, $\alpha_0=10^4, 10^5$, (b) $\gamma=0.4$, $\alpha_0=6.25, 62.5$, (c) $\gamma=1$, $\alpha_0=1, 10$. The end points of the curves $r_i(\beta)$ in (a) correspond to the LO instability.}
	\label{fig:Fig14}
\end{figure} 

Shown in Fig. \ref{fig:Fig14} is $r_i(\beta)$ calculated at $l=10\lambda$, $\zeta_n=0.8$, $n_i=0.25 \lambda^{-3}$, and different values of $\gamma$ and $\alpha_0$.  At the lowest frequency $\gamma=0.01$ and $\alpha_0=10^4$ the elastic skin depth  $L_\omega=\lambda/\sqrt{2\pi\gamma}\approx 4\lambda$ is about half of the vortex length and the surface resistance decreases with field due to the decrease of the vortex viscosity with $v$. This mechanism is similar to that for a vortex pinned by a strong single defect ~\cite{Manula}.

\begin{figure}[h!]
		\centering
		\includegraphics[width=\columnwidth]{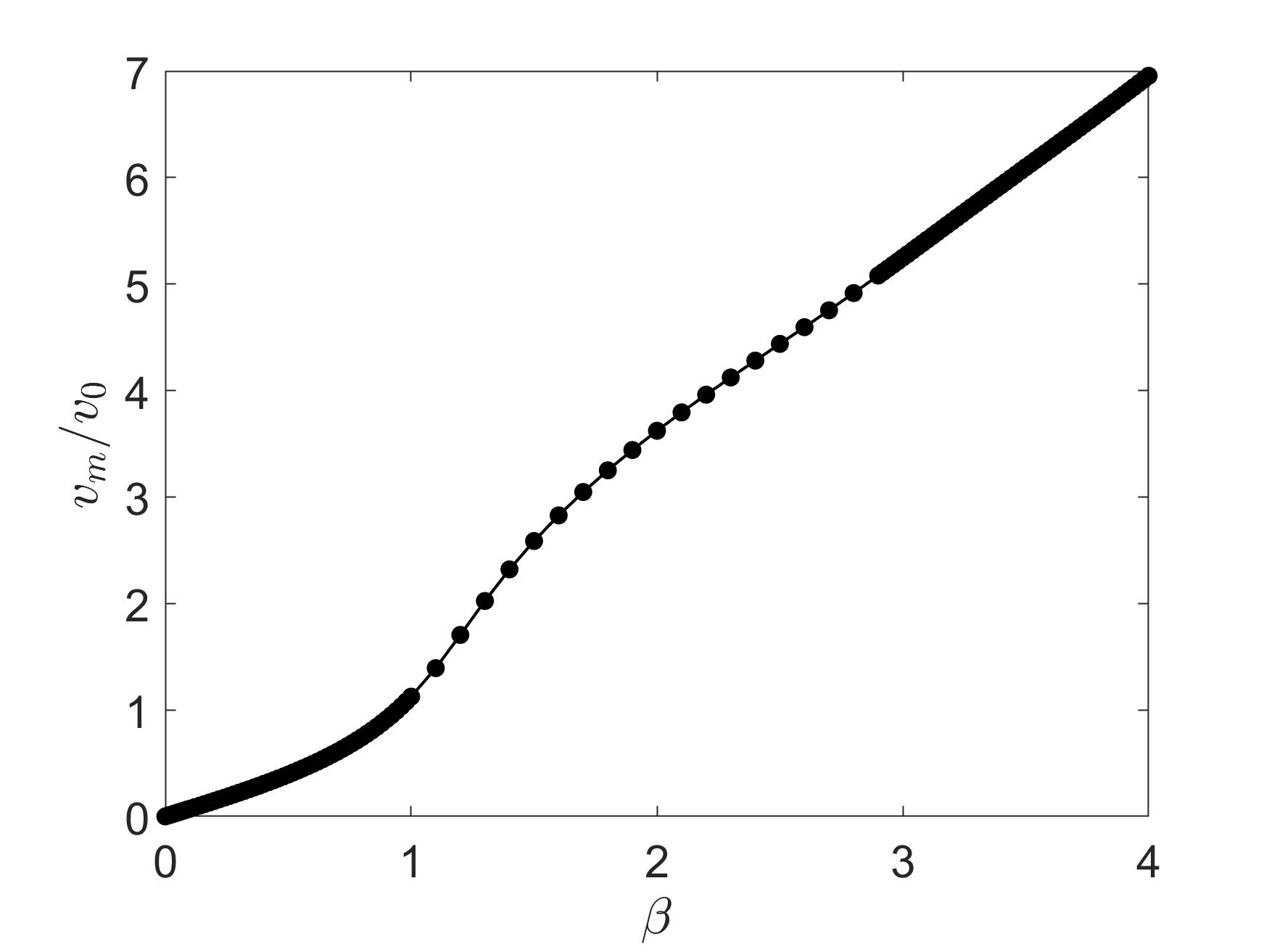}
		\caption{The peak velocity of the vortex tip $v_m=v_0\sqrt{\alpha}\left(\dot{u}_y^2+\dot{u}_z^2\right)^{1/2}$ calculated at  
		$\alpha_0=\gamma=1$.}
		\label{fig:Fig15}
	\end{figure}

The behavior of $r_i(\beta)$ changes at higher $\omega$, as shown in Figs.  \ref{fig:Fig14}b and c. Here $r_i(\beta)$ becomes nonmonotonic, the peaks in $r_i(\beta)$ shifting to lower fields as the LO parameter $\alpha_0=(\lambda f_0/v_0)^2$ increases.  At the peak of $r_i(\beta)$ at $\beta=\beta_m$, the maximum velocity of the vortex tip $v_m$ becomes of the order of $v_0$, but no LO jumps occur because of the restoring effect of the vortex line tension. The increase of $r_i(\beta)$ with $\beta$ at $\beta<\beta_m$ is mostly due to the increase of $L_\omega\sim[\epsilon/\eta(v)\omega]^{1/2}$ caused by the decreasing $\eta(v)$. At $\beta\gtrsim\beta_m$ the elastic skin depth  $L_\omega$ becomes of the order of the vortex length and $r_i(\beta)$ decreases with $\beta$ due to the decrease of $\eta(v)$ with $v$~ \cite{Manula}. The field dependence of the peak velocity $v_m$ of the vortex tip is shown in Fig. \ref{fig:Fig15}.

As $\beta$ exceeds $\beta_m$, the dynamics of the vortex tip changes from a nearly harmonic oscillations at $\beta\lesssim \beta_m$ to highly anharmonic oscillations at $\beta\gtrsim \beta_m$, the amplitude of oscillations increasing greatly at $\beta\gtrsim \beta_m$  ~\cite{SI}.  Bending distortions along the vortex are mostly confined within the elastic skin depth  $L_\omega$ which increases with $v$ due to LO reduction of $\eta(v)$ and eventually becomes larger than $l$ at $v_m\gtrsim v_0$.  Our simulations also show that the nonlinear dynamics of the vortex at $\beta>\beta_m$ becomes dependent on the vortex mass. Here the peaks in $r_i(\beta) $ shift to higher $\beta$ as the frequency increases and a larger Lorentz force is required to accelerate the vortex above $v_0$.

\begin{figure}[h!]
	\centering
	\includegraphics[width=\columnwidth]{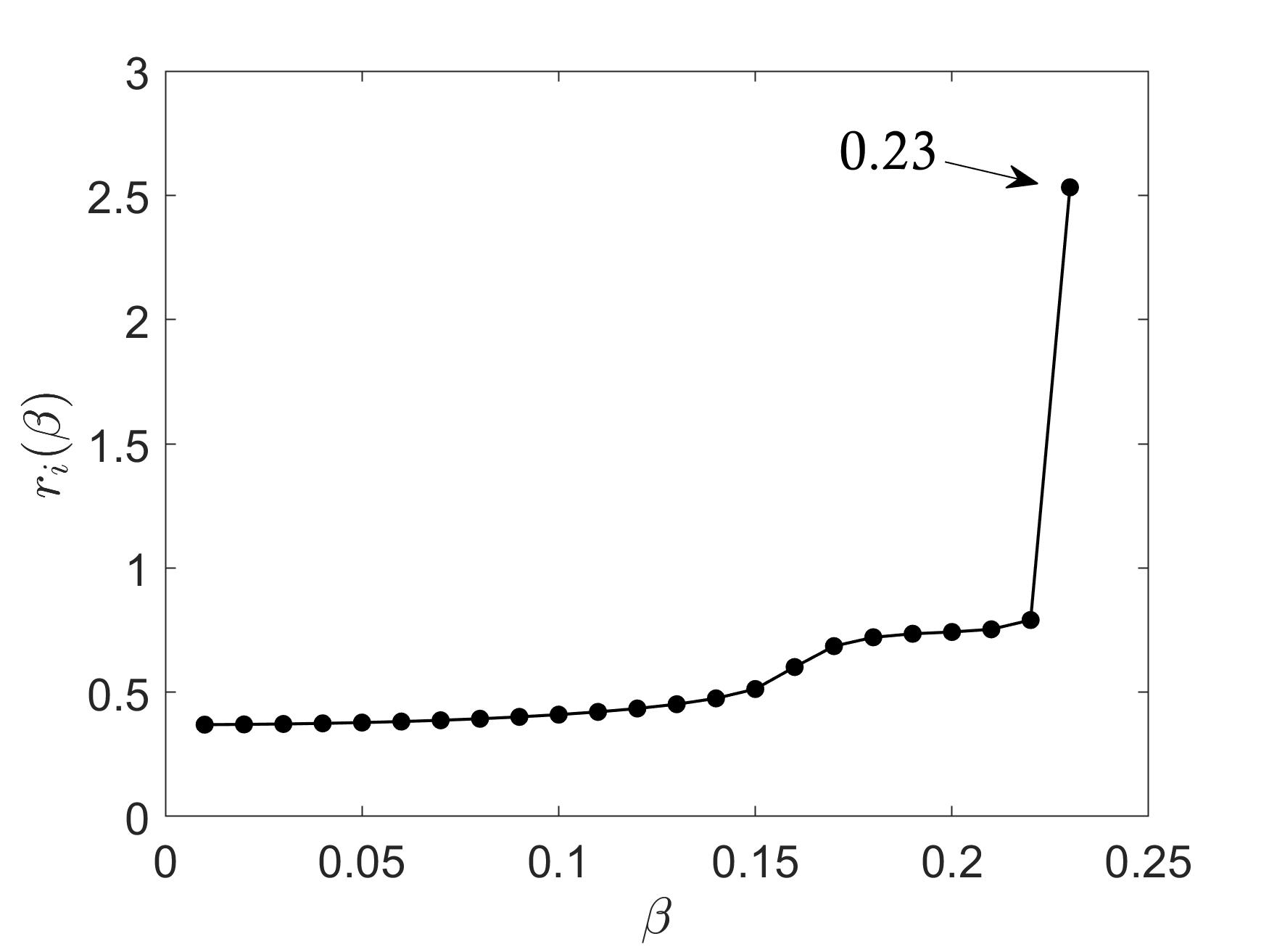}
	\caption{$r_i(\beta)$ calculated for bulk pinning at $l/\lambda=10$, $\kappa=2$, $\zeta_n=0.8$, $ n_i=0.25 \lambda^{-3} $, $\gamma=0.1$, and $\alpha_0=100$. The arrow shows the onset of the LO instability.}
	\label{fig:Fig16}
\end{figure} 

It turns out that neither the vortex line tension nor pinning can always suppress the LO instability if $\alpha_0$ is large enough. For instance, Fig. \ref{fig:Fig16} shows $r_i(\beta)$ calculated at $\gamma=0.1$, $\alpha=\alpha_0\gamma^2=1$, and the same pinning and materials parameters as in Fig. \ref{fig:Fig14}c.  The dynamics of the vortex was simulated by slowly ramping up $\beta(t)=0.01t$. At $\beta=\beta_m\approx 0.16$, the velocity of the vortex tip $v_m$ becomes of the order of $v_0$ and a hump in $r_i(\beta)$ develops. As $\beta$ further increases, the bending oscillations along the vortex extend all the way through the film thickness, giving rise to the LO instability of the entire vortex at $\beta>0.23$ marked by the arrow in Fig. \ref{fig:Fig16}. Details of the vortex dynamics and movies of the LO instability of a vortex in a film in the presence of bulk pinning are given Ref. \onlinecite{SI}.

\begin{figure}[h!] 
	\centering
	\includegraphics[width=\columnwidth]{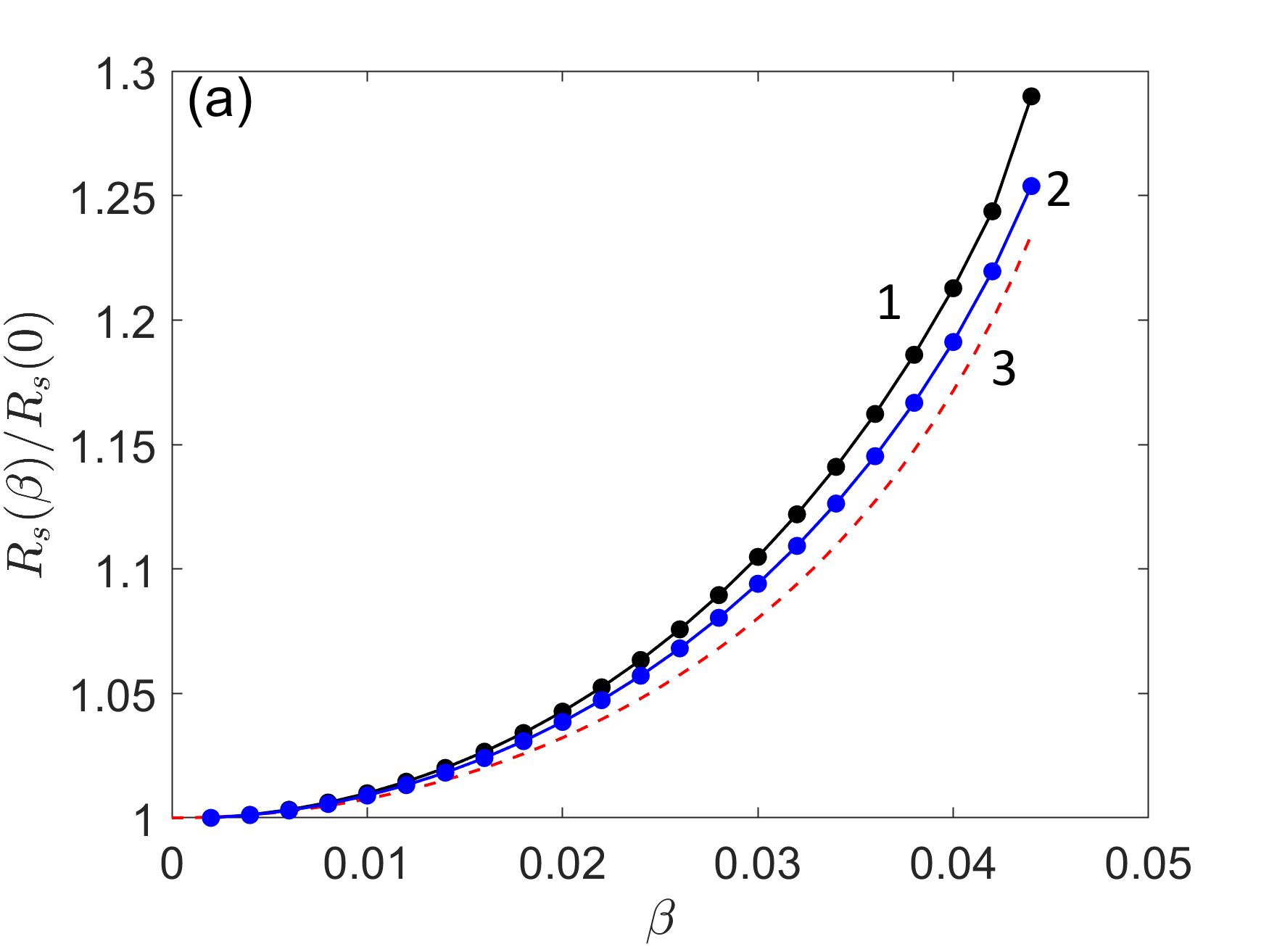}
	\includegraphics[width=\columnwidth]{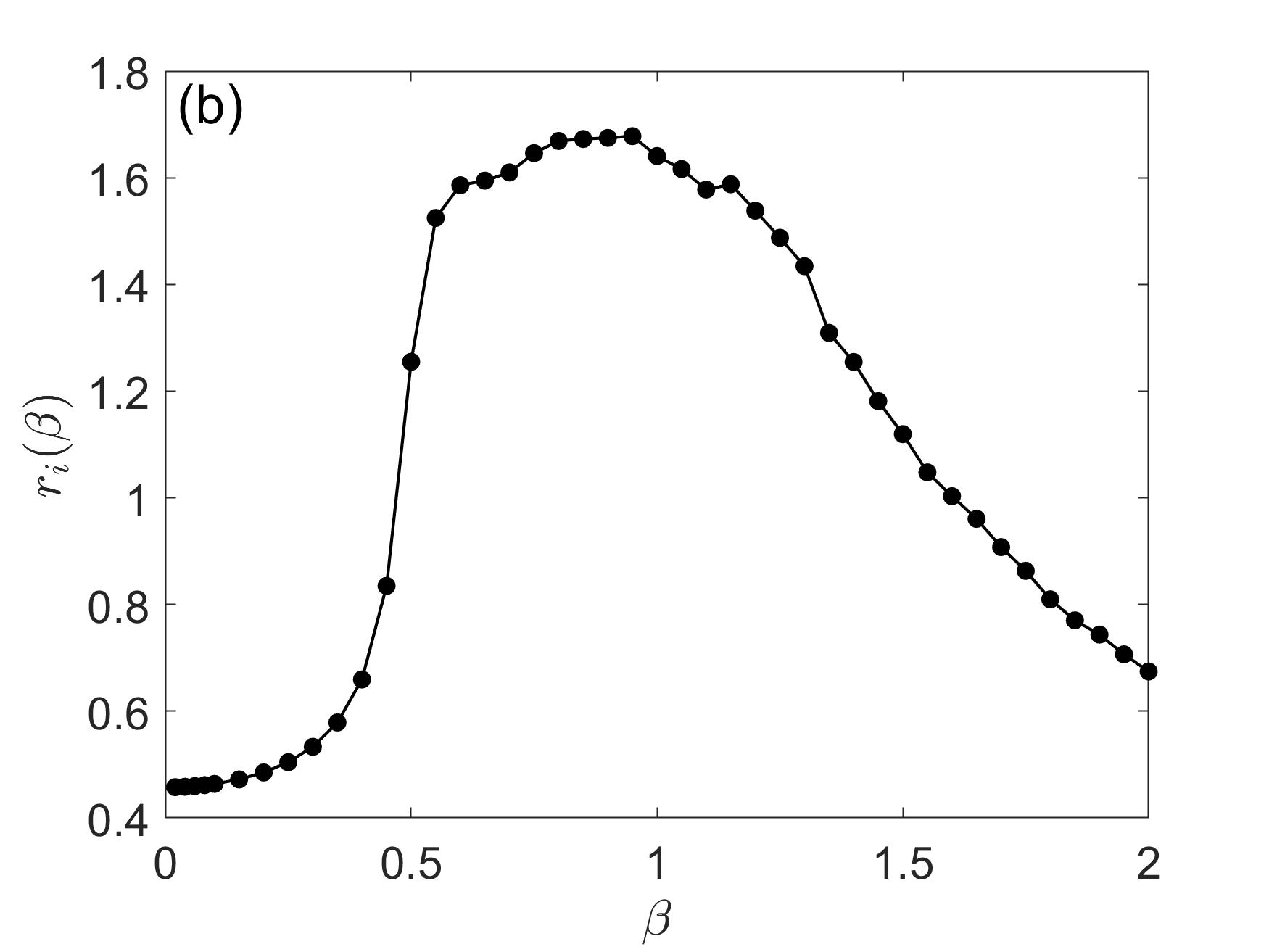}
	\includegraphics[width=\columnwidth]{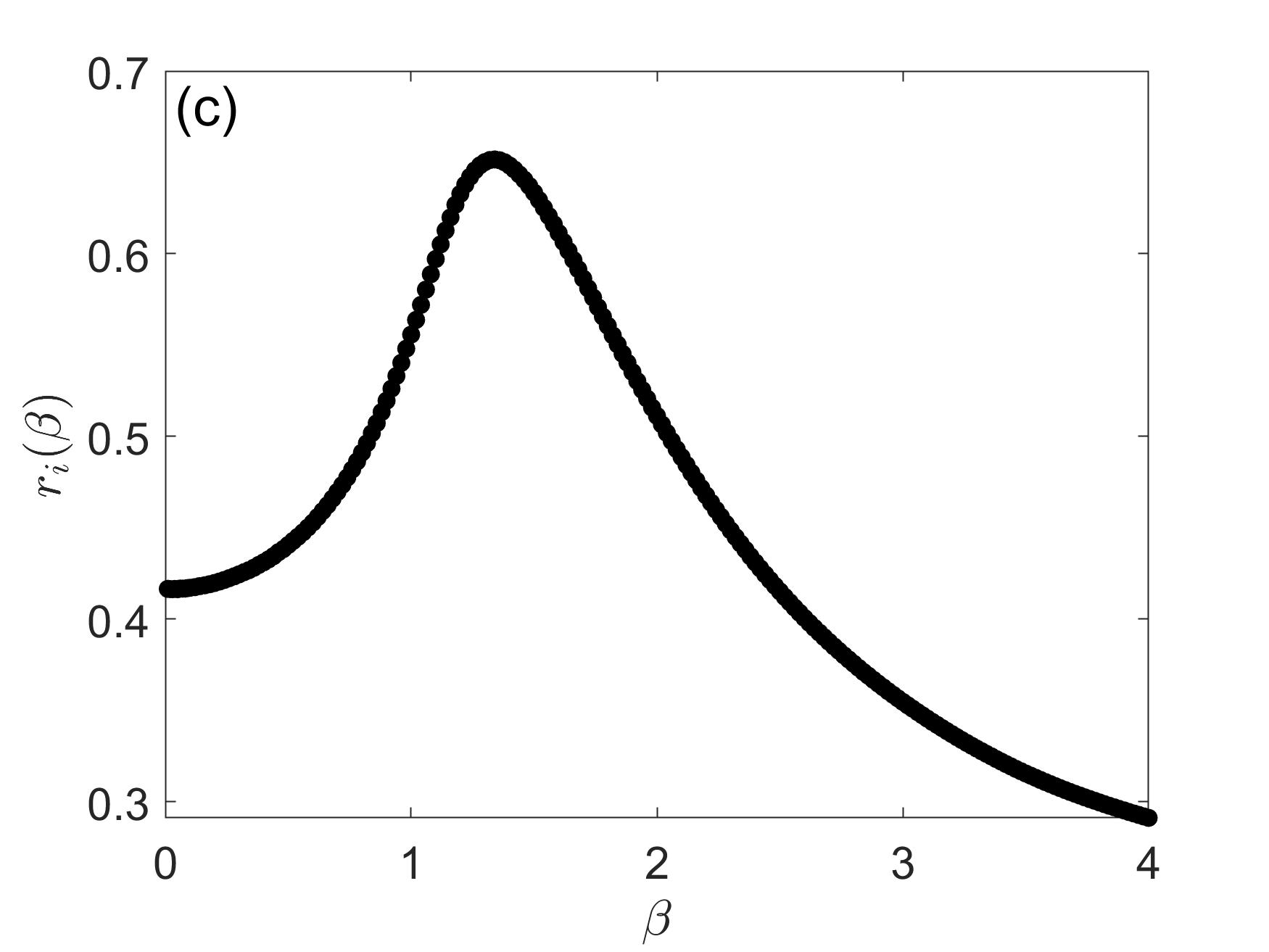}
	\caption{$r_i(\beta) $ calculated at $l/\lambda=10$, $\kappa=2$ for bulk pinning with the pin density $n_i=0.25 \lambda^{-3}$ and: (a) 
	$\gamma=0.01$, $\alpha_0=10^4$, $\zeta_n=0.08$ (1), $\zeta_n=0.04$ (2) and the dashed line given by Eqs. (\ref{riw}) and (\ref{bet1}) with $\beta_1=2\beta\alpha_0^{1/2}\lambda/l=20\beta$, (b) $\gamma=0.4$, $\alpha_0=6.25$, $\zeta_n=0.08$ (c) $\gamma=1$, $\alpha_0=1$, $\zeta_n=0.08$. The end points of the curves $r_i(\beta)$ in (a) correspond to the LO instability.}
	\label{fig:Fig17}
\end{figure} 

The surface resistance calculated for weaker pinning ($\zeta_n=0.08$) at different values of $\gamma$ and $\alpha_0$ is shown in Fig. \ref{fig:Fig17}. The behavior of $r_i(\beta)$ at higher frequencies ($\gamma \gtrsim 0.1$) is qualitatively similar to that is shown in Fig. \ref{fig:Fig14}b and c for $\zeta_n=0.8$, as $r_i(\beta)$ is mostly controlled by the vortex drag and elasticity.  Yet at low frequencies $r_i(\beta)$ for weaker pinning increases with $\beta$ while $r_i(\beta)$ for stronger pinning decreases with $\beta$. These different behaviors of $r_i(\beta)$ at $\gamma\ll 1$ could be understood as follows.  

At $\gamma=0.01$, we have $L_\omega\simeq l/2$ so the vortex oscillates as a nearly straight stick.  For stronger pinning, the velocity of the vortex is mostly determined by the balance of the Lorentz and pinning forces. As a result, the LO instability is mitigated by pinning and $r_i(\beta)$ decreases with $\beta$ due to the decrease of $\eta(v)$ with $v$ all the way to $v\gg v_0$. For weaker pinning, $v(t)$ is mostly controlled by the balance of the Lorentz and drag forces, so $v(\beta)$ can only increase up to $\simeq v_0$. In this case $R_i$ can be calculated from the force balance for a straight vortex in a pinning-free film of thickness $l>\lambda$:
\begin{equation}
\frac{l\eta_0 v}{1+(v/v_0)^2}=\phi_0H(t).
\label{b1}
\end{equation}
Hence, $v(t)=v_0\tilde{\beta}/[1+(1-\tilde{\beta}^2)^{1/2}]$,
where $\tilde{\beta}(t)=\beta_1\sin\omega t$ and $\beta_1=2H\phi_0/\eta_0v_0l$.  From the mean power $\bar{p}=H\phi_0 l\langle v(t)\sin \omega t\rangle$, we obtain $R_i=2\bar{p}n_\square/H^2$ in the form:
\begin{gather}
\frac{R_i(H)}{R_i(0)}=\frac{4}{\pi}\int_0^\pi\frac{\sin^2 tdt}{1+(1-\beta_1^2\sin^2t)^{1/2}},
\label{riw} \\
 \beta_1=\frac{2H\phi_0}{\eta_0v_0l}=2\beta\alpha_0^{1/2}\frac{\lambda}{l},
 \label{bet1}
\end{gather}
where $R_i(0)=\phi_0^2n_\square/\eta_0 l$ and $n_\square$ is the mean density of trapped vortices.  According to Eq. (\ref{riw}), $r_i(\beta)$ increases with $\beta$ as shown by the dashed line in Fig. \ref{fig:Fig17}a. The model captures the behavior of $r_i(\beta)$ calculated from  Eqs. (\ref{dyneq1}) and (\ref{dyneq2}), the agreement between the model and the numerical results improves as $\zeta_n$ decreases. Here $\beta=\beta_1$ is the onset of the LO instability of the vortex in a film.

\section{RF overheating} \label{Heat}
In the this section we address the effects caused by overheating of fast vortices. We first show that overheating of a single vortex can produce the LO-like velocity dependence of $\eta(v)$ at low frequencies and a field dependence of $\eta(H)$ at high frequencies. Then we consider how the global rf overheating of arrays of trapped vortices can affect the field dependence of the surface resistance. 

\subsection{Overheating of a single vortex. }

Consider a straight perpendicular vortex moving with a velocity $v(t)$ in a film of thickness $l<L_\omega$ at high fields $H\gg H_p$ for which pinning is negligible. The temperature $T({\bf r},t)$ around a vortex is described by a thermal diffusion equation ~\cite{gm}:
    \begin{equation}
    \nu\partial_tT=\nabla\cdot (k\nabla T) - W(T)+\eta_0(T_m)v^2f[y-u(t),z].
    \label{tdeq}
    \end{equation}
Here $\nabla$ is a 2d gradient in the $yz$ plane parallel to the film surface, $\nu(T)$ is the specific heat, $k(T)$ is
the thermal conductivity, $W(T)$ describes the heat exchange between the film and the substrate, and the last term in Eq. (\ref{tdeq}) accounts for the power produced by the vortex core moving with the velocity $v(t)=du/dt$. To separate the effect of overheating from the LO mechanism, we disregard here the LO velocity dependence of the isothermal $\eta(v)$ and consider the Bardeen-Stephen $\eta_0(T_m)$ taken at the core temperature $T_m[v(t)]$, where a bell-shaped function $f(r)$ with $\int f({\bf r})d^2{\bf r}=1$ accounts for a finite core size. For qualitative estimates, we  take $f(r)=(\pi\xi^2)^{-1}\exp (-r^2/\xi^2)$ but the details of $f(r)$ affect weakly the results presented below. 

We consider the case of $\omega\tau_\epsilon\lesssim 1$ in which quasiparticles and phonons have the same temperature $T$ and both contribute to $\nu(T)$ and $k(T)$ in Eq. (\ref{tdeq}). Then $W(T) = (T-T_0)\alpha_K/l$, where $\alpha_K(T)$ is the Kapitza thermal conductance caused by acoustic mismatch of phonons scattered at the interface between the film and a He bath or a substrate, and $T_0$ is the bath temperature \cite{gm,He}. This approximation of $W(T)$ is valid if the film is thin enough $lk<\alpha_K$ so that $T$ is nearly constant across the film thickness ~\cite{gm}. For weak overheating $T_m-T_0 \lesssim T_0 $, the parameters $\nu(T)$, $k(T)$ and $\alpha_K(T)$ can be taken at $T=T_0$ and Eq. (\ref{tdeq}) brings the following length, time and velocity scales of heat diffusion in a film:
\begin{equation}
L_\theta=\sqrt{\frac{k l}{\alpha_K}},\quad t_\theta=\frac{\nu l}{\alpha_K}, \quad v_\theta=\frac{L_\theta}{t_\theta}.
\label{thermu}
\end{equation}
Here $L_\theta$ is a lateral decay length of $T(r)$ and $t_\theta$ is a cool down time of the film. The condition $lk<\alpha_K$ of a nearly constant $T$ along the film thickness implies $L_\theta\gg l$. For a $0.5\,\mu$m thick Nb$_3$Sn film with $k=10^{-2}$ W/mK, $\nu=10^2$ J/m$^3$K, and $\alpha_K=2.5$ kW/m$^{2}$K at 2K~ \cite{He,cody}, we get $L_\theta\simeq 1.4\,\mu$m, $t_\theta=20$ ns and $v_\theta=70$ m/s.  

Solution of the linearized Eq. (\ref{tdeq}) gives an integral equation for $T_m(t)$ obtained in Appendix C. This integral equations can be solved analytically in two limiting cases of $\omega t_\theta\ll 1$ and $\omega t_\theta\gg 1$. For a slowly-varying $v(t)$ at $\omega t_\theta\ll 1$, the normalized temperature of the vortex core $\theta_m(t)=(T_m(t)-T_0)/(T_c-T_0)$ is determined by its instantaneous velocity:
\begin{gather}
\theta_m(t)=\frac{v^2(t)}{v_0^2+v^2(t)},
\label{tetm} \\
v_0^2=\frac{2\pi k\rho_nT_c}{\phi_0 B_{c2}(0)\cal{L}}, \quad {\cal L}=\ln\frac{L_\theta}{\tilde{\xi}}-\frac{1}{2}\ln\left(1+\frac{v^2}{4v_\theta^2}\right),  
\label{vot}
\end{gather}  
where $\tilde{\xi}=\xi e^{\gamma_E/2}/2\approx 0.67\xi$ and $\gamma_E=0.577$. Substituting $\theta_m$ from Eq. (\ref{tetm}) into 
$\eta(T_m)=(1-\theta_m)\eta_0$ yields the LO velocity dependence (Eq. \ref{LO}) of the non-isothermal $\eta_0(\theta_m)$. The weak logarithmic dependence of $v_0$ on $v$ in Eq. (\ref{vot})  does not prevent the vortex runaway at $v\gtrsim v_0$ as the drag force cannot balance the Lorentz force. For a $0.5\,\mu$m thick Nb$_3$Sn film with $\xi_0\simeq 2.8$ nm, $L_\theta/\tilde{\xi}\simeq 750$, $T_c=16$ K, $\rho_n=0.36\,\mu\Omega$m, and $B_{c2}(0)\simeq 30$ T ~\cite{hc2}, Eq. (\ref{vot}) gives $v_0\simeq 1.15$ km/s at $v=v_0$. 

The temperature $\theta(y,z,t)=(T-T_0)/(T_c-T_0)$ around a moving vortex at $\omega t_\theta\ll 1$ is given by: 
\begin{gather}
\theta(y,z,t)=\frac{v^2}{(v_0^2+v^2){\cal L}}e^{-\frac{v(y-u(t))}{2D}}K_0\left(\frac{R}{L_\theta}\right),
\label{thetm} \\
R=\left[((y-u(t))^2+z^2+\xi_1^2)(1+v^2/4v_\theta^2)\right]^{1/2},
\label{Ruu} 
\end{gather}
where $u(t)=\int v(t)dt$ is the core position along $y$, $K_0(x)$ is a modified Bessel function, and $\xi_1\approx 3\xi/4$. At $v\ll v_\theta$ Eq. (\ref{thetm}) describes a moving hotspot in which $\theta(y,z,t)$ decreases exponentially over the thermal length $L_\theta$. At $v\gg v_\theta$  the decay length of $\theta(y,z)$ along $z$ shrinks to $L_\perp=L_\theta(1+v^2/4v_\theta^2)^{-1/2}\simeq 2D_\theta/v\ll L_\theta$, while the shape of $\theta(y,z)$ along the direction of motion $y$ becomes a saw-tooth like, with a sharp front of width 
$L_f=L_\theta /(v/2v_\theta+\sqrt{1+v^2/4v_\theta^2})\simeq D_\theta/v\ll L_\theta$ and a trailing tail of length $L_b=L_\theta(v/2v_\theta+\sqrt{1+v^2/4v_\theta^2})\simeq vt_\theta\gg L_\theta$. 

At high frequencies $\omega t_\theta\gg 1$, the temperature field cannot follow the oscillating vortex which produces a nearly stationary hotspot in which $v(t)$ is controlled by a field-dependent non-isothermal viscosity $\eta_0(\theta_m)$. Here $\theta_m$ in the center of the hotspot is determined by the mean power $\bar{p}=\langle v^2(t)\eta(T_m)f(y-u(t))\rangle$ to be calculated self-consistently from Eq. (\ref{tdeq}).  As shown in Appendix C, $\theta_m(B)$ is determined by the equation: 
\begin{gather}
\theta_m(1-\theta_m)=b^2\ln\left[\Xi(1-\theta_m)/b\right],
\label{tmm}\\
b=\frac{B}{B_\theta},\qquad B_\theta=\frac{l\mu_0}{\xi_0}\left(\frac{2kT_c}{\rho_n}\right)^{1/2}\!\left(1-\frac{T_0}{T_c}\right),
\label{ba} \\
\Xi=\frac{3.7\tilde{L}_\theta}{u_0},\qquad \tilde{L}^2_\theta=\frac{lk}{\alpha_K-H^2R_T/2}.
\label{Xii} 
\end{gather}  
Here $B=\mu_0H$, $u_0=\rho_nB_\theta/B_{c2}(0)\mu_0l\omega$, and $B_\theta$ is a thermal field scale. The term $R_TH^2/2$ in $\tilde{L}_\theta$, which takes into account quasiparticle heating, is essential for a global thermal runaway considered below, where $R_T=\partial R_{BCS}/\partial T$, and $R_{BCS}$ is the BCS surface resistance~\cite{tf1}.  Equation (\ref{tmm}) is a power balance of the lateral heat and the power from the vortex, $4\pi (T_m-T_0)k\sim \eta_0(T_m)v_m^2\ln (\tilde{L}_\theta/u_m)$, where $v_m\simeq B\phi_0/\mu_0\eta_0(\theta_m)l$ and $u_m=v_m/\omega$ are the amplitudes of vortex velocity and displacement, respectively. The factor $\ln(\tilde{L}_\theta/u_m)$ accounts for a reduction of $\theta_m$ due to spreading the power over a larger area as $u_m$ increases.      

From the mean sheet power $\bar{P}=n_\square\phi_0^2H^2/2l\eta_0(\theta_m)l$ generated by vortices of areal density $n_\square=B_0/\phi_0$, we obtain the surface resistance $R_i=B_0\phi_0/l\eta_0(\theta_m)$:
\begin{equation}
R_i=\frac{\rho_n B_0 }{(1-\theta_m)lB_{c2}(0)}=\frac{R_i(0)}{1-\theta_m},
\label{Rio}
\end{equation}  
For a Nb$_3$Sn film with $k=10^{-2}$ W/mK, $\rho_n=0.36\,\mu\Omega$m, $T_c=16$ K,  $B_{c2}(0)\simeq 30$ T ~\cite{hc2}, $l=0.5\, \mu$m, $L_\theta=1.4\, \mu$m, we obtain $B_\theta\simeq 185$ mT and $\Xi\approx 11$ at $1.2$ GHz.

\begin{figure}[h!] 
	\centering
	\includegraphics[scale=0.5]{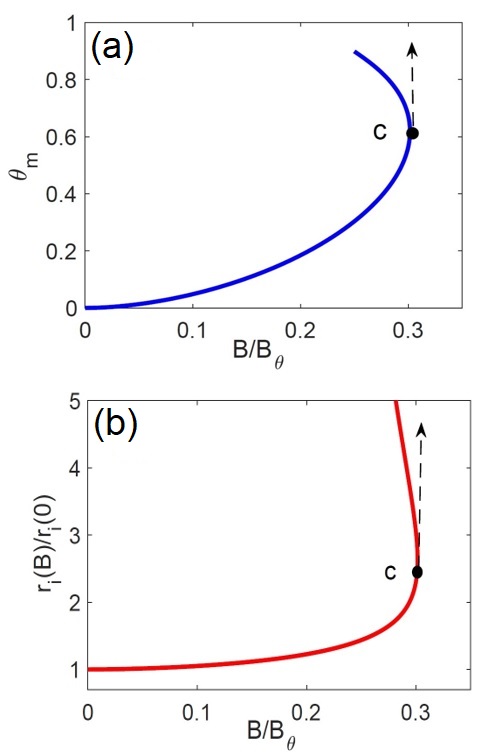}
	\caption{(a) The overheating $\theta_m(B)$ and (b) the surface resistance $r_i(B)$ calculated from Eqs. (\ref{tmm})-(\ref{Rio}) for the parameters of Nb$_3$Sn given in the text and $R_TH^2\ll \alpha_K$. The point c marks the onset of a runaway vortex instability. Upper branches of $\theta_m(B)$ and $r_i(B)$ correspond to unstable states. }
	\label{fig:Fig18}
\end{figure} 

Shown in Fig. (\ref{fig:Fig18}) are $\theta_\theta(B)$ and $R_i(B)$ calculated from Eqs. (\ref{tmm})-(\ref{Rio}). Both $\theta_m(B)$ and $R_i(B)$ are multivalued functions with lower and upper branches corresponding to stable and unstable states, respectively. As $B$ increases, 
$\theta_m(B)$ increases until $B=B_t\simeq 0.3 B_\theta$ at point $c$, where $\theta_m(B_t)=\theta_{mc}\simeq 0.6$. 
Here $B_t$ is a field threshold of instability of an oscillating vortex.  

The critical velocity of the vortex $v_0$ at $B=B_t$ can be evaluated from the force balance $B_t\phi_0/\mu_0=\eta_0(1-\theta_{mc})lv_0$, 
which yields $v_0=\rho_n B_t/\mu_0 l(1-\theta_{mc})B_{c2}(T_0)$. Using $B_\theta$ from Eq. (\ref{ba}), we present $v_0$ in the form:
\begin{equation}
v_0=\frac{C\rho_nB_\theta}{\mu_0lB_{c2}(T_0)}=C\left(\frac{4\pi k\rho_nT_c}{\phi_0B_{c2}(0)}\right)^{1/2},
\label{v0a}
\end{equation}
where $C=B_t/(1-\theta_{mc})B_\theta$. For the case shown in Fig. \ref{fig:Fig18}, $C\simeq 3/4$ and $v_0\simeq 2.6$ km/s.

The above results show that at $\omega t_\theta\ll 1$, the overheating of a single vortex produces the LO-like dependence of the non-isothermal Bardeen-Stephen $\eta_0(\theta_m)$. At $\omega t_\theta\gg 1$, the vortex oscillates in a self-induced hotspot and $\eta_0(\theta_m)$ is controlled by the field amplitude. In both cases $v_0$ is determined by the same combination of the materials parameters except for the different logarithmic factors ($\ln(v/v_\theta)$ in Eq. (\ref{vot}) and $\ln(\tilde{L}_\theta/u_m)$ in (\ref{v0a})):
\begin{equation}
v_0\propto \xi_0\sqrt{k\rho_n T_c} 
\label{vou}
\end{equation}
For Nb$_3$Sn at 4K, Eq. (\ref{vot}) and (\ref{v0a}) give $v_0\simeq 1-3$ km/s which depends logarithmically on the film thickness and $\omega$. 
These values of $v_0$ are about 2-6 times larger than $v_0$ for the LO mechanism estimated above from Eq. (\ref{diff}). Overheating could 
facilitate the LO instability by reducing the effective critical velocity $v_0$ determined by both heat and quasiparticle diffusion. At the same time,  overheating would decrease the energy relaxation time $\tau_\epsilon(T)$ in Eqs. (\ref{LO})-(\ref{taup}) and thus reduce the contribution of the LO mechanism to $\eta(v)$ at $T\ll T_c$.    

In the models presented here the overheating of the vortex core can reach $\theta_m\simeq 0.5-0.6$ at $v=v_0$, which would reduce the vortex line tension and pinning energies in Eq. (\ref{eq1}).  At $\omega t_\theta\gg 1$, this effect can be taken into account by field-dependent  superconducting parameters in Eqs. (\ref{eq1}) and (\ref{eq2}) at the core temperature $T_m(H)$ to be calculated self-consistently from Eq. (\ref{tdeq}).
The linearized Eq. (\ref{tdeq}) which neglects the temperature dependencies of the parameters is applicable qualitatively at $T_m-T_0\lesssim T_0$. If $T_0\ll T_c$, the temperature dependencies of the quasiparticle thermal conductivity $k\simeq k_n(\Delta/k_BT)^2\exp(-\Delta/k_BT)$ ~\cite{cody}, the lattice specific heat $\nu\propto T^3$ and the heat flux from the film, $W(T)\propto T^4-T_0^4$ should be taken into account. At $T\ll T_c$ the phonon thermal conductivity can exceed the quasiparticle contribution ~\cite{cody}, but cooling a hot vortex by diffusion of phonons is ineffective if the phonon thermal wavelength $\lambdabar_T \simeq 2\pi\hbar c_s/k_BT$ exceeds the core size $\xi$. For instance, $\lambdabar_T\simeq 55$ nm at 4.2 K is well above $\xi_0\simeq $ 3nm in Nb$_3$Sn.  If $v_0$ is mostly limited by the quasiparticle $k(T)$, Eq. (\ref{vou}) gives $v_0\propto (T_c/T)D^{1/2}\exp(-\Delta_0/2k_BT)$, where we used the Wiedemann-Frantz law $\rho_nk_n\sim(k_B/e)^2T_c$. In this case the dependencies of $v_0$ on $T$ and the mean free path are similar to those of the LO critical velocity (\ref{vol}). Moreover, $v_0(T)$ can decrease exponentially with $T$ 
both in the LO and in the overheating model, if $\tau_\epsilon$ in Eq. (\ref{vol}) is determined by 
recombination of quasiparticles ~\cite{lo9}. Because of slow electron-phonon energy relaxation rate  $\Gamma \propto T_e^5\exp(-\Delta/k_BT_e)$ due to quasiparticle recombination ~\cite{relax}, the temperature of quasiparticles $T_e$ around moving vortices can be higher than the lattice temperature ~\cite{shklovsk}.  

The inhomogeneities of $T({\bf r},t)$ around moving vortices diminish strongly if trapped vortices are 
spaced by distances shorter than $ L_\theta$ and the vortex hotspots overlap. In this case the thermal runaway  
instability of fast vortices is controlled by the strong temperature dependence of the quasiparticle surface resistance 
$R_{BCS}(T)$, as it is shown in the next subsection. This instability occurs at a rather weak global overheating 
$T-T_0\simeq k_BT_0^2/\Delta_0$.          

\subsection{Global RF overheating}

We now turn to the global overheating produced by trapped vortices with the flux density $B_0\ll B_{c1}$. If vortices are spaced by distances $\lesssim L_\theta$, the hotspots around vortices overlap and $T(H)$ at the surface exposed to the rf field is nearly uniform. Then $T(H)$ is determined by the following equation of a thermal feedback model ~\cite{gurevich2012}:
\begin{equation}
\left[\bar{R}_i(T,H)+R_{\rm BCS}(T)\right]\frac{H^2}{2}=\frac{(T-T_0)k\alpha_K}{k+l\alpha_K}.
\label{tb}
\end{equation}
This equation represents a balance of the rf power generated at the surface of a slab of thickness $l$ and the heat flux going across the slab to the coolant (or a substrate) at the ambient temperature $T_0$. The total surface resistance in Eq. (\ref{tb}) is a sum of the  BCS quasiparticle contribution $R_{\rm BCS}$ and the vortex residual resistance $\bar{R_i}$ averaged over the surface.  
At $T\ll T_c$ and $\hbar\omega\ll\Delta$ the low-field $R_{\rm BCS}(T)$ in the dirty limit is given by ~\cite{ag_sust,gk}:  
\begin{equation}
R_{\rm BCS}(T)  
\simeq \frac{\mu_0^2\omega^2\lambda^3\Delta}{\rho_nk_BT} \ln \biggl( \frac{9k_B T}{4\hbar\omega} \biggr) e^{-\Delta/k_BT}.
\label{eq:Rs_MB}
\end{equation}
For weak overheating, $R_{\rm BCS}(T)\simeq R_{\rm BCS}(T_0)\exp[(T-T_0)\Delta_0/k_BT_0^2]$, and dependencies of $R_i$, $\alpha_K$ and $k$ on $T$ can be neglected. Then Eq. (\ref{tb}) reduces to:
\begin{gather}
\left[h\frac{r_i(\beta)}{r_i(0)}+e^{a\theta_0}\right]\beta^2=s\theta_0,
\label{tb1} \\
h=\frac{R_i(0)}{R_{\rm BCS}},\qquad s=\frac{2T_0\mu_0^2k\alpha_K}{(k+l\alpha_K)R_{\rm BCS}B_{c1}^2},
\label{tb2}
\end{gather}
where $\theta_0=(T-T_c)/T_0$, $a=\Delta_0/k_BT_0$, $h$ is proportional to the mean density of trapped vortices, and all parameters in $h$ and $s$ are taken at $T=T_0$. For Nb cavities screened to a few $\%$ of the Earth's magnetic field, $h$ is typically around $0.1-0.5$ and raises to $\gtrsim 10$ for unscreened cavities at $2$K and 1-2 GHz ~\cite{Padam_book,gigi} The solution of Eq. (\ref{tb1}) determines a non-isothermal surface resistance $R_s(\beta)$ including both the BCS and the vortex contributions:
\begin{equation}
R_s(\beta)=\left[h\frac{r_i(\beta)}{r_i(0)}+e^{a\theta_0}\right]R_{\rm BCS}.
\label{tb3}
\end{equation} 
For Nb with $l=3$ mm, $k=7$  W/mK, $\alpha_{K}=2.5$ kW/m$^{2}$K, $R_{\rm BCS}=20$ n$\Omega$, $\kappa=2$,  $B_{c1}\approx $ 84 mT,  $\lambda=60$ nm, $\Delta_0=1.8k_BT_c$, and $f_0=37$ GHz at 1.5 K, we obtain $s = 43$, $a=11$, and $\gamma=0.04$ at $f\approx 1.5$ GHz. For Nb$_3$Sn with $R_{\rm BCS}(T_{0})\simeq 20$ n$\Omega$ at $4.2$ K, $\lambda=111$ nm, $B_{c1}\simeq $ 49 mT, $k\simeq10^{-2}$ W/mK, $\Delta_0=1.9k_BT_c$, we have $a\approx 8$, $f_0=175$ GHz and $s\simeq 395$ at $l=3\,\mu$m.

\begin{figure}[ht]
	\centering
	\includegraphics[width=\columnwidth]{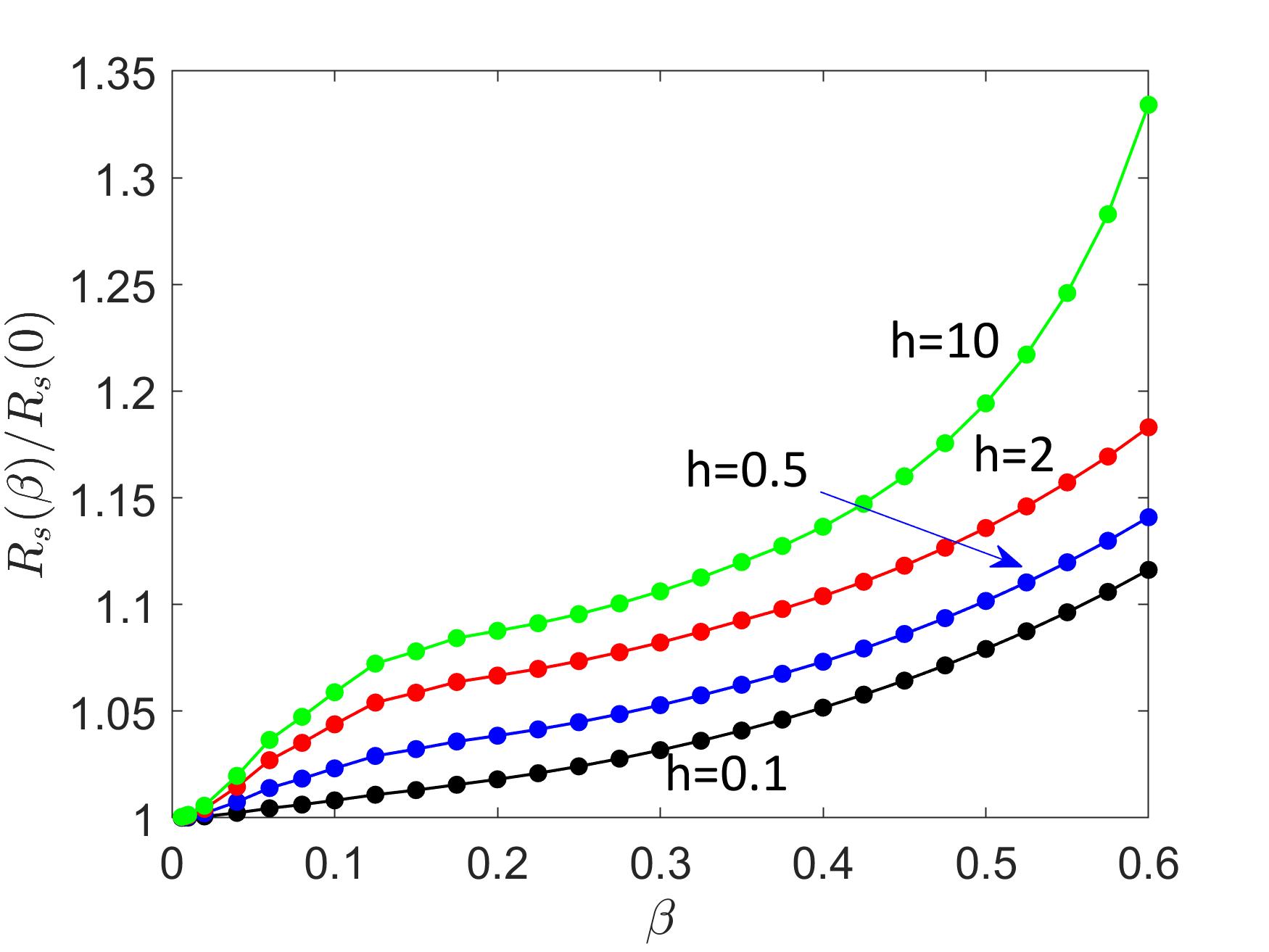}
	\caption{Non-isothermal $R_s(\beta)$ calculated from Eqs. (\ref{tb1}) and (\ref{tb3}) at different flux density parameters $h$, $				\gamma=0.04$, $s=43$, $a=11$ and the isothermal 
	$R_i(\beta)$ taken from Fig. \ref{fig:Fig3}b.}
	\label{fig:Fig19}
\end{figure} 

A non-isothermal $R_s(H)$ calculated from Eqs. (\ref{tb1}) and (\ref{tb3}) at low fields for which the velocity dependence of $\eta(v)$ is negligible, is shown in Fig. \ref{fig:Fig19}, where the isothermal vortex contribution $R_i(\beta)$ at $\gamma=0.04$ is taken from Fig. \ref{fig:Fig3}b.  Heating causes an upturn of $R_s(H)$ at higher fields, masking the saturation of $R_i(H)$ shown in Fig. \ref{fig:Fig3}b. 
This can be understood by expanding Eq. (\ref{tb1}) in $a\theta_0\lesssim 1$ at low fields, which yields $\theta_0= [1+hr_i(\beta)/r_i(0)]\beta^2/(s-a\beta^2)$, where $a\beta^2$ in the denominator is retained because $a\gg 1$. Then,
\begin{equation}
R_s(\beta)=\frac{R_{BCS}+R_i(\beta)}{1-a\beta^2/s}.
\label{rst}
\end{equation} 
The upturn in $R_s(\beta)$ comes from the decrease of the denominator in Eq. (\ref{rst}) with $\beta$, which describes the effect of thermal feedback on $R_{BCS}(T)$ at low fields ~\cite{gc,tf2}.  The thermal balance cannot be sustained above the breakdown field $B_b$ obtained from the condition $\partial \beta/\partial\theta_0=0$, where $\beta(\theta_0)$ is defined by Eq. (\ref{tb1}).  For instance, $\beta_b=B_b/B_{c1}=(s/ae)^{1/2}$ at $h=0$ is reached at a critical weak overheating  $\theta_b=k_BT_0/\Delta_0\ll 1$ ~\cite{gurevich2012}.  

\begin{figure}[ht]
	\centering
	\includegraphics[width=\columnwidth]{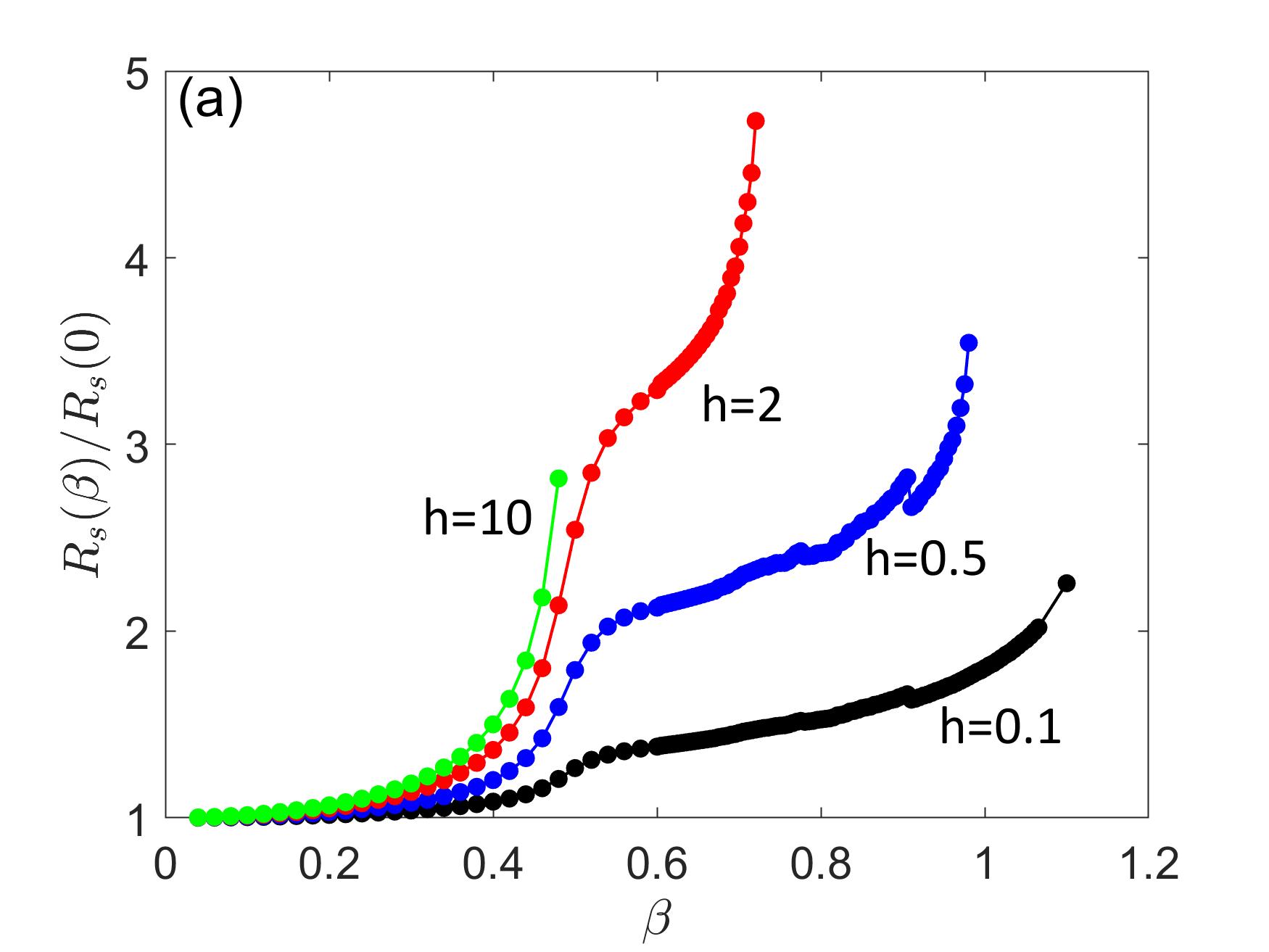}
	\includegraphics[width=\columnwidth]{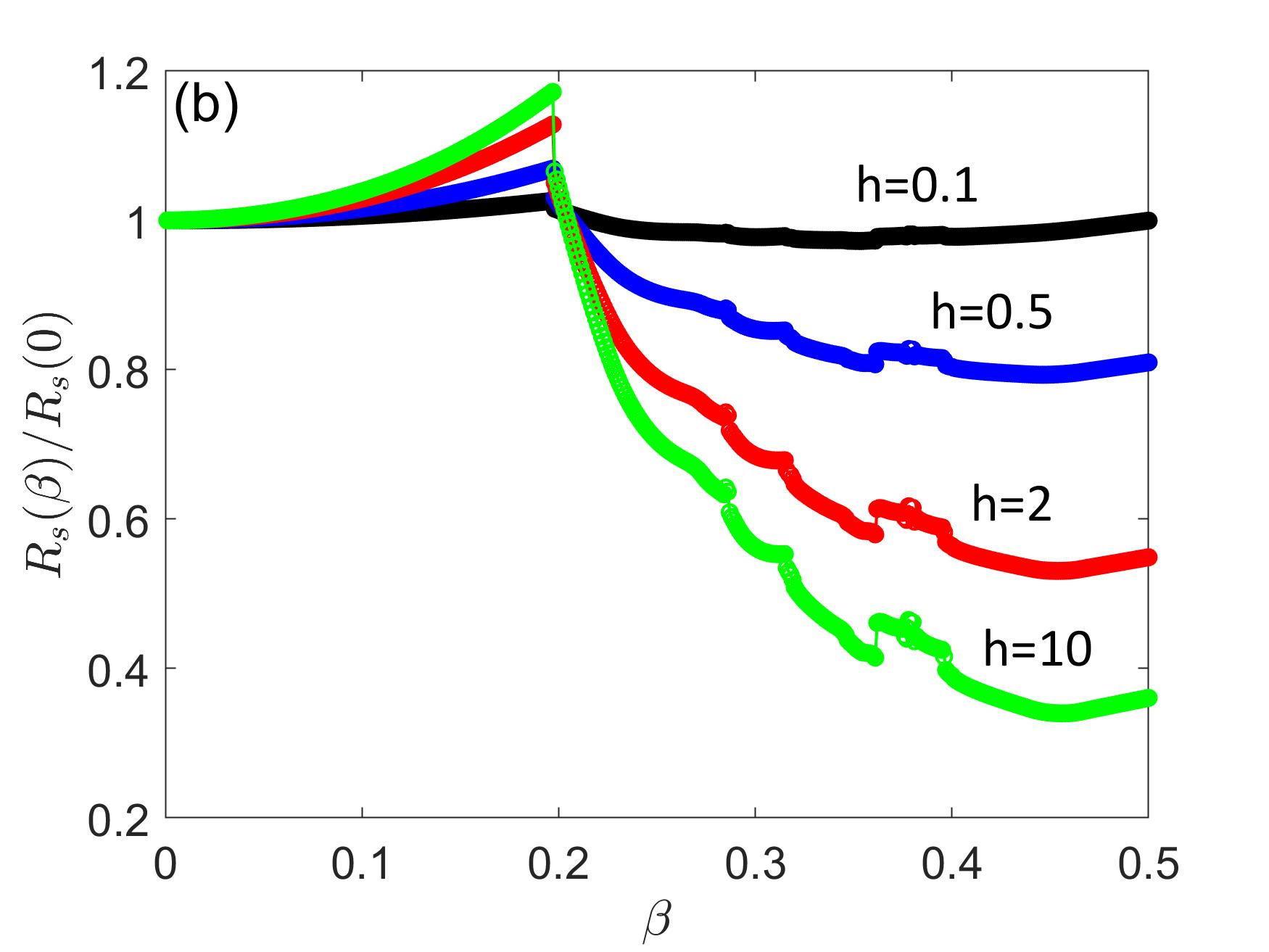}
	\caption{Non-isothermal $R_s(\beta)$ calculated from Eqs. (\ref{tb1}) and (\ref{tb3}) for: (a) $\gamma=0.4$, $\alpha_0=6.25$; (b) $\gamma=0.04$, 		$\alpha_0=100$. In both cases, $\zeta_n=0.8$, $s=43$ and $a=11$. The end points of the curves $R_s(\beta)$ correspond to the thermal 
	breakdown in (a) and to the LO instability in (b). }
	\label{fig:Fig20}
\end{figure} 

Now we turn to the effect of heating on $R_s(H)$ for the LO velocity-dependent vortex drag.
Figure \ref{fig:Fig20}a shows $R_s(\beta)$ for different trapped flux densities $\propto h$ calculated  
at $\alpha_0=6.25$, $\gamma=0.4$ and other parameters specified in the caption, assuming that $v_0$ is independent of $B_0$.  Here heating masks the descending field dependence of the isothermal $R_i(\beta)$ and results in $R_s(\beta)$ raising with $\beta$.  The end points of $R_s(\beta)$ correspond to the thermal breakdown field $B_b$ which in this case is smaller than the field onset of the LO instability. 

If the parameter $\alpha_0=(\lambda f_0/v_0)^2$ is large enough, the LO instability can occur at a field smaller than $B_b$ and heating may not fully mask the descending $R_i(H)$. For instance, Fig. \ref{fig:Fig20}b shows $R_s$ calculated at $\alpha_0=100$ and $\gamma=0.04$. Here the descending field dependencies of $R_s(H)$ persist at all $h$ up to the end points of the curves $R_s(\beta)$. In this case the end points  correspond to the LO instability at a field $\beta\approx 0.5$ smaller than the thermal breakdown field $\beta_b$.

\section{Discussion}

Sparse trapped vortices moving through a random pinning landscape can produce nonsystematic field dependencies of the local surface resistance $R_i(H)$ controlled by different pin configurations with the same density $n_i$ in a volume $\sim \lambda^2L_\omega$. Yet the global  $\bar{R}_i(H)$ obtained by averaging the local $R_i(H)$ over different pin configurations, increases smoothly with $H$ at small fields and levels off at higher fields.  In addition, $\bar{R}_i(H)$ increases linearly with the frequency at small $\omega$, indicating that the rf losses mostly come from hysteretic depinning of the vortex from multiple pins, unlike $R_i(\omega)\propto \omega^2$ for small-amplitude vortex oscillations~ \cite{Gittleman}.  

As $H$ increases, the velocity dependence of the vortex drag can result in a nonmonotonic field dependence of $\bar{R}_i(H)$ if the tip of a trapped vortex moves faster than the LO critical velocity.  Bending rigidity of the vortex and pinning suppress the LO instability up to a certain field, yet we observed vortex jumps in thick films at higher fields, even in the case of strong pinning. Understanding rf losses of  trapped vortices in films is important for  increasing the quality factors and rf breakdown field by multilayer coating ~\cite{ml1,ml2,ml3,ml4} or deposition of thick 
$(l\gg\lambda)$ Nb$_3$Sn films onto the inner surface of Nb cavities ~\cite{film1,film2}.  We did not observe a dynamic shape instability of the vortex due to the LO decrease of $\eta(v)$, which requires taking into account nonlinear elasticity of the vortex ~\cite{Manula}.   

Measurements of $R_s(H,T)$ at low $T$ may offer an opportunity to extract the LO critical velocity $v_0(T)$ at $T\ll T_c$ because global 
heating effects for sparse vortices driven by Meissner currents are weaker than in magneto-transport measurements ~\cite{lo1,lo2,lo3,lo4,lo5,lo6,lo7,lo8,lo9,lo10,lo11}. Yet the overheating of a single vortex can also contribute to the LO-like decrease of $\eta(v)$ with $v$ but mask the decrease of $R_s(H)$ with $H$ for higher densities of trapped vortices $B_0/\phi_0$. The effect of density of trapped vortices on $R_i(H)$ is  complex. On the one hand, the LO velocity $v_0$ decreases with $B_0$ at $B_0\gtrsim \phi_0/L_d^2$ and local peaks of $T({\bf r},t)$ around vortices diminish at $B_0\gtrsim \phi_0/L_\theta^2$, which shifts the descending part of $R_i(H)$ to lower fields. On the other hand, the global overheating increases with $B_0$, resulting in smaller breakdown fields, as shown in  Fig. \ref{fig:Fig20}a. This problem can be mitigated by measuring $R_i(H)$ at lower frequencies.

The microwave reduction of $R_i(H)$ resulting from $\eta(v)$ decreasing with $v$ can contribute to the decrease of $R_s(H)$ with $H$ observed on Nb resonators ~\cite{cav1,cav2,cav3,cav4,cav5,cav6}.  Other contributions to this effect could come from a nonlinear quasiparticle conductivity \cite{ag_prl} which can be tuned by magnetic impurities and proximity-coupled oxide layer at the surface \cite{kg,gk}. The vortex contribution can be separated from other contributions by measuring $R_s$ at different frequencies and densities of trapped flux. Given the extreme sensitivity of high-Q resonant cavities, the relevant densities of trapped vortices may be achieved by varying the degree of screening of the Earth's magnetic field.      

\section{ACKNOWLEDGMENTS}
This work was supported by NSF under Grants PHY 100614-010 and PHY 1734075, and by 
DOE under Grant DE-SC 100387-020.

\appendix
%%%%%%%%%
\section{Dissipated power}  \label{Ap1}

At $M=0$ the dynamics of a vortex in a pinning potential $U({\bf r})$ is determined by the relaxation equation:
\begin{equation}
\eta\frac{\partial \bold{u}}{\partial t}=-\frac{\delta G}{\delta \bold{u}},
\label{rela}
\end{equation}
where $G\{{\bf u}\}=\int_0^l \varrho dx$ is a free energy functional, and 
\begin{gather}
\varrho=\frac{\epsilon}{2}\left(\frac{\partial u_z}{\partial x}\right)^2\!+\frac{\epsilon}{2}\left(\frac{\partial u_y}{\partial x}\right)^2\!+
%\nonumber \\
U(x,{\bf u}) -yF_L.
\label{G_u}
\end{gather}
Hereafter we use ${\bf u}$, $x$ and $l$ in natural length units, unless noted otherwise. 
Eqs. (\ref{rela}) and (\ref{G_u}) give:
\begin{gather}
\eta \partial_t u_y=\epsilon \partial_{xx}u_y-\partial_y U + F_L,
\label{dy1}\\
\eta\partial_t u_z=\epsilon \partial_{xx}u_z -\partial_z U.
	\label{dy2}
\end{gather}
Multiplying Eqs. (\ref{dy1}) and (\ref{dy2}) by $v_y=\partial_t u_y$ and $v_z=\partial_t u_z$, adding them and integrating over $x$ and $t$, 
yields:  
\begin{gather}
\int_0^l \langle \eta v^2-{\bf v}\cdot{\bf F}_p-{\bf v}\cdot{\bf F}_L\rangle dx =
\nonumber \\
\epsilon\int_0^l\langle\partial_{xx}u_y\partial_t u_y+\partial_{xx}u_z\partial_t u_z\rangle dx,
\label{int}
\end{gather}
where ${\bf F}_p=-\nabla U$ is the pinning force. Next, we expand  
$u_y(x,t)$ and $u_z(x,t)$ in the Fourier series:
\begin{gather}
u_y=\sum_n [A_y(x,n)\cos n\omega t+B_y(x,n)\sin n\omega t],
\label{Y} \\
u_z=\sum_n [A_z(x,n)\cos n\omega t+B_z(x,n)\sin n\omega t].
\label{Z}
\end{gather} 
Substituting Eqs. (\ref{Y}) and (\ref{Z}) in the r.h.s. of Eq. (\ref{int}), we observe that all terms proportional to $A^2$ and $B^2$ vanish after time averaging giving:
\begin{gather}
\frac{\omega}{2}\int_0^l\sum_nn[B_y\partial_{xx}A_y-A_y\partial_{xx}B_y + 
\nonumber \\
B_z\partial_{xx}A_z-A_z\partial_{xx}B_z]dx
\label{av}
\end{gather}
After integration by parts Eq. (\ref{av}) becomes:
\begin{equation}
\frac{\omega}{2}\sum_n n\bigl[B_y\partial_x A_y -A_y\partial_x B_y+ B_z\partial_x A_z -A_z\partial_x B_z\bigr]_0^l.
\label{parts}
\end{equation}
Because of the boundary condition $\partial_x u_y=\partial_x u_z=0$ at the surface or $u_y=u_z=0$ at a strong pin,
the expression in the brackets of Eq. (\ref{parts}) vanishes. Thus,
\begin{equation}
P=\int_0^l \langle \eta v^2-{\bf v}\cdot{\bf F}_p\rangle dX=\int_0^l\langle {\bf v}\cdot{\bf F}_L\rangle dx.
\label{pu}
\end{equation}
The power results from the work against the viscous and pinning forces, including hysteretic ${\bf F}_p$ for strong pinning.

\section{Labusch criterion for a single vortex.}  \label{Ap2}

\begin{figure}[h!]
	\centering
	\includegraphics[width=\columnwidth]{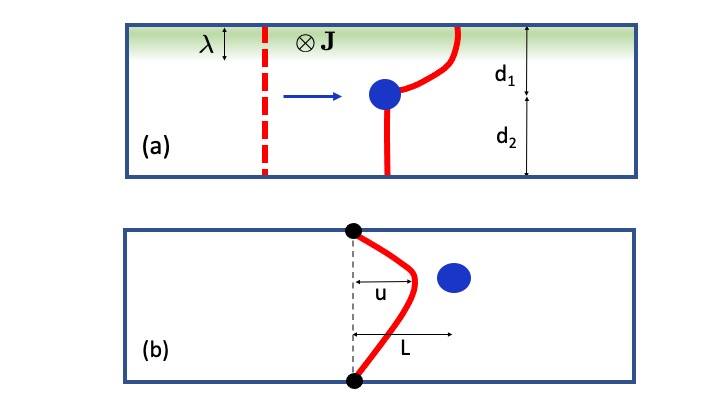}
	\caption{(a) A vortex driven toward the pin (blue circle). Dashed and solid lines show the initial and final shapes of the free and pinned vortex, 			respectively; 
	(b)  a vortex with fixed ends interacting with a pinning center}
	\label{fig:Fig21}
\end{figure}

Consider the vortex driven toward the pin by the Meissner current in a film, as shown in Fig. \ref{fig:Fig21}. The vortex gets depinned if the Lorentz force $\phi_0 H$ exceeds the pinning force. This force balance for a single vortex in a film does not involve the interplay of vortex pinning and bending distortion which can cause a hysteretic pinning response of the vortex in an infinite FLL which remains undisturbed far away from the pin \cite{sp1,sp2,sp3,sp4,sp5}. The latter implies that the ends of a long vortex are fixed in their equilibrium positions in the ideal FLL, unlike the free ends of a vortex in the film shown in  Fig. \ref{fig:Fig21}a.      
 
The situation changes if the ends of the vortex interacting with a pin inside the film are fixed by surface defects, as depicted in  Fig. \ref{fig:Fig21}b. For the short-range core pinning, the balance of elastic and pinning forces yields the following equation for the displacement $u$: 
\begin{equation}
\left(\frac{u}{\sqrt{u^2+d_1^2}}+\frac{u}{\sqrt{u^2+d_2^2}}\right)\epsilon=f_p(L-u), 
\label{fb1}
\end{equation}
where $f_p(y)=-\int_{-\infty}^\infty \partial_y U(x,y,0)dx$ and $U(x,y,0)$ is defined by Eq. (\ref{eq3}).
For weak bending distortions $u\ll (d_1,d_2)$, Eq. (\ref{fb1}) can be recast to:
\begin{gather}
L=\frac{gu_1}{\left[1+(u_1/\xi)^2\right]^{3/2}}+u_1,
\label{L} \\
g=\frac{\pi U_nd_1d_2}{\epsilon\xi(d_1+d_2)}=\frac{\pi\zeta_n d_1d_2}{2\kappa\lambda(d_1+d_2)},
\label{defL}
\end{gather}
where $u_1=L-u$ is the minimum distance between the pin and the distorted vortex (see  Fig. \ref{fig:Fig21}b). Here $g$ is proportional to the dimensionless pinning parameter $\zeta_n=2\kappa^2U_n/\epsilon$ and depends on the pin position.

\begin{figure}[h]
	\centering
	\includegraphics[scale=0.35]{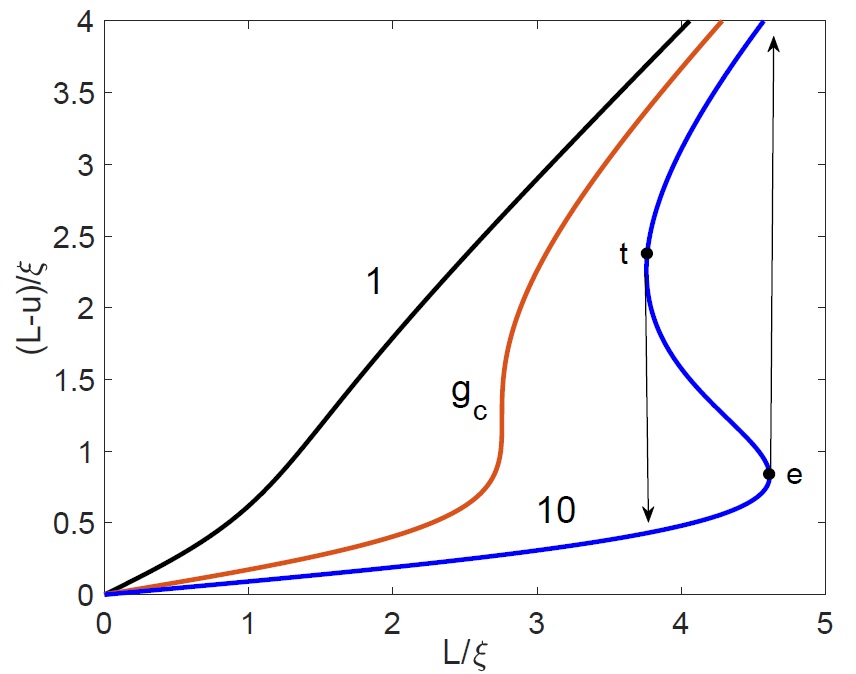}
	\caption{Dependencies of the minimum pin-vortex spacing $u_i=L-u$ on $L$. The labels at the curves denote the values of  	
		$g$, where $g_c=(5/2)^{5/2}/2\approx 4.94$.}
	\label{fig:Fig22}
\end{figure}

Shown in Fig. \ref{fig:Fig22} is $u_1(L)$ calculated from Eq. (\ref{L}). For weak pining $(g\lesssim 1)$, the vortex remains nearly straight so $u_1(L)\approx L$. As $g$ increases, the vortex bows out toward the pin and $u_1(L)$ becomes a multivalued function at $g>g_c=(5/2)^{5/2}/2\approx 4.94$, where $g_c$ is obtained from $dL/du_1=d^2L/du_1^2=0$. At $g>g_c$ the pinning response of the vortex is hysteretic. 

The change in the behavior of $u_1(L)$ upon increasing the pinning force shown in  Fig. \ref{fig:Fig22}, appears similar to that of a vortex in an infinite FLL ~  \cite{sp1,sp2,sp3,sp4,sp5}, the condition $g>g_c$ being analogous to the Labusch criterion \cite{sp1}. Yet, unlike the Labusch parameter which quantifies the strength of a single pin and depends on the vortex spacing in the FLL, the parameter $g$ depends on the film thickness and the pin position, so it is not a characteristic of the pin strength only. For instance, the vortex with fixed ends can have a non-hysteretic pinning response if the pin is near the surface, where $g<g_c$ and $d_1\ll d_2$, but exhibit a hysteretic pinning response if the same pin is inside the film, where $g>g_c$ at $d_1\simeq d_2$. Pinning response of a single vortex in a film could become hysteretic only if several pins are involved.  For identical, randomly-distributed pins assumed in our simulations, the vortex can adjust its position and straighten up to mitigate the hysteretic pinning response. This may extend the applicability of the 2D collective pinning theory to $l> L_c$, as it is evident from Fig. 6.  Thus, the classification of weak and strong pinning centers based on the Labusch criterion for a vortex in an infinite FLL is not applicable to a single vortex in a film.

\section{Temperature of a moving vortex} \label{Ap3}

Eq. (\ref{tdeq}) can be written in the dimensionless form:
    \begin{equation}
    \dot{\theta}=\nabla^2\theta - \theta+q(\theta_m(t),t)f[y-u(t),z],
    \label{a1}
    \end{equation}
where $q(\theta_m,t)=\eta_0(\theta_m)v^2(t)/(T_c-T_0)k$,  $\theta=(T-T_0)/(T_c-T_0)$, $\theta_m(t)=(T_m-T_0)/(T_c-T_0)$ and
time and coordinates are normalized by the respective thermal scales (\ref{thermu}).  For the Bardeen-Stephen temperature dependence of $\eta_0(T)=\eta_0(0)(1-T/T_c)$, we have $\eta_0(\theta_m)=(1-\theta_m)\eta_0$, where $\eta_0$ implies $\eta_0(T_0)$. 

Solving Eq. (\ref{a1}) by the Fourier transform ~\cite{gc} results in a differential equation for
$\theta_{\bf p}(t)=\int \theta({\bf r},t)\exp(-i{\bf pr})d^2{\bf r}$:
    \begin{equation}
    \dot{\theta}_{\bf p}+(1+p^2)\theta_{\bf p}=f_{\bf p}q(t)e^{-ip_yu(t)},
    \label{a2}
    \end{equation}
If the vortex starts moving at $t=-\infty$, the solution of Eq.  (\ref{a2}) is:
    \begin{equation}
    \theta_{\bf p}(t)=f_{\bf p}\int_0^\infty e^{-(1+p^2)t'-ip_yu(t-t')}q(t-t')dt'.
    \label{a3}
    \end{equation}
For $f(r)=(\pi\xi^2)^{-1}\exp(-r^2/\xi^2)$ and $f_{\bf p}=\exp(-p^2\xi^2/4)$, the inverse Fourier transform of Eq. (\ref{a3})
yields (in original units):
    \begin{equation}
    \theta({\bf r},t)=\frac{1}{\pi}\int_0^\infty\frac{dt'q(t-t')}{4t'+t_0}
    e^{-\frac{t'}{t_\theta}-\frac{[y-u(t-t')]^2+z^2}{(t_0+4t')D_\theta}},
    \label{a4}
    \end{equation}
where $t_0=\xi^2/D_\theta \ll t_\theta$, and $D_\theta=k/\nu$. Setting $y=u(t)$, $z=0$ in Eq. (\ref{a4}) gives an 
integral equation for $\theta_m(t)$ at the moving vortex core:
  \begin{gather} 
    \theta_m(t)=\frac{\eta_{0}}{\pi(T_c-T_0) k}\int_0^\infty\frac{v^2(t-t')[1-\theta_m(t-t')]}{4t'+t_0}\times \nonumber \\
    \exp\left[-\frac{t'}{t_\theta} -\frac{[u(t)-u(t-t')]^2}{(4t'+t_0)D_\theta}\right]dt',
    \label{tm}
    \end{gather}
    This equation is supplemented by an equation for the velocity of a straight vortex of length $l$:
    \begin{equation}
    l\eta_0(1-\theta_m)v=\phi_0H\sin\omega t
    \label{dynv}
    \end{equation}

At $\omega t_\theta\lesssim (1+v^2/v_\theta^2)^{1/2}$ the integrand in Eq. (\ref{tm}) 
decreases exponentially over $t'\sim t_\theta(1+v^2/4v_\theta^2)^{-1/2}$. In this case the slowly varying $v(t-t')$ and $\theta_m(t-t')$ can be replaced 
with $v(t)$, $\theta_m(t)$, and $u(t)-u(t-t')\approx v(t)t'$. Then integration in Eq. (\ref{a4}) at $t_0\to 0$ yields:
\begin{gather}
\theta=\frac{\eta(\theta_m)v^2}{2\pi k(T_c-T_0)}e^{-\frac{v(y-u)}{2D_\theta}}K_0\left(\frac{R}{L_\theta}\right),\quad R\gtrsim \xi,
\label{temr} \\
R=\left[((y-u)^2+z^2)(1+v^2/4v_\theta^2)\right]^{1/2},
\label{RR}
\end{gather}
where $u=\int v(t)dt$, and $K_0(x)$ is a modified Bessel function. 
A logarithmic singularity in $\theta(y,z,t)$ in Eq. (\ref{temr}) at $R\to 0$ is cut off by taking into account 
$t_0=\xi^2/D_\theta$ in Eqs. (\ref{a4}) and (\ref{tm}) \cite{gc}. For slowly-varying $v(t)$ and $\theta_m(t)$, integration in Eq. (\ref{tm}) produces an integral exponential function ~\cite{abr}, giving at $t_0\ll t_\theta$:
\begin{equation}
\!\!\theta_m=\frac{\eta_0v^2(1-\theta_m)}{4\pi(T_c-T_0) k}\left[2\ln\frac{2L_\theta}{\xi}-\gamma_E-\ln\!\left(1+\frac{v^2}{4v_\theta^2}\right)\right].
\label{tmv}
\end{equation}
This equation leads to Eqs. (\ref{tetm})-(\ref{vot}).

At $\omega t_\theta\gg 1$ the vortex oscillating with the amplitude  $u_m\simeq \phi_0H/l\eta_0\omega\ll L_\theta$ produces a  
nearly stationary hotspot, in which   
$\theta(y,z,t)$ is close to $\bar{\theta}(x,z)$ averaged over the rf period. Here $\bar{\theta}(x,z)$ satisfies the stationary thermal diffusion equation:
\begin{gather}
\!\!\!\nabla^2\bar{\theta}-\frac{\bar{\theta}}{\tilde{L}_\theta^2}+\frac{\eta_0(\theta_m)}{(T_c-T_0)k}\langle v^2(t)\delta(y-u(t))\rangle\delta(z)=0,
\label{steq} \\
\tilde{L}^2_\theta=lk/[\alpha_K-H^2R_T/2].
\label{Ltt}
\end{gather}  
Here $f(y,z)$ is replaced with $\delta(y-u(t))\delta(z)$ at $\xi\ll u_m\ll L_\theta$, and $R_TH^2/2$ in Eq. (\ref{Ltt}) accounts for quasiparticle heating, where $R_T=\partial R_{BCS}/\partial T$ ~\cite{tf1}. 
Using $v(t)=v_m\sin\omega t$ and $u(t)=-u_m\cos\omega t$ with $v_m=\phi_0H/l\eta_0(\theta_m)$ and $u_m=v_m/\omega$, we obtain the solution of Eq. (\ref{steq}):
\begin{gather}
\!\!\!\!\bar{\theta}=\frac{2b^2}{1-\theta_m}\langle K_0\biggl[\frac{\sqrt{(y+u_m\cos\omega t)^2+z^2}}{\tilde{L}_\theta}\biggr]\!\sin^2\omega t\rangle, 
\label{tav} \\
b=\frac{B}{B_\theta},\qquad B_\theta=\frac{l\mu_0}{\xi_0}\sqrt{\frac{2kT_ck}{\rho_n}}\left(1-\frac{T_0}{T_c}\right),
\label{ta}
\end{gather}
where $\eta_0=\phi_0^2/2\pi\xi_0^2\rho_n$ was used. The self-consistency equation for $\theta_m$ in the center of the hotspot is obtained by setting $\bar{\theta}(x,y)=\theta_m$ at $y=z =0$ in Eq. (\ref{ta}):
\begin{equation}
\theta_m(1-\theta_m)=2b^2\langle K_0[|\cos\omega t|u_m/\tilde{L}_\theta]\sin^2\omega t\rangle. 
\label{tam} 
\end{equation}

Using $K_0(x)=\ln (2/x)-\gamma_E$ ~\cite{abr}  at $u_m\ll L_\theta$, we calculate the average $I=\langle ...\rangle = (\omega/2\pi)\int_0^{2\pi/\omega}... dt$ in Eq. (\ref{tam}):
\begin{equation}
I=\frac{2}{\pi}\int_0^{\pi/2}\!\!dt\left[\ln\left(\frac{2\tilde{L}_\theta}{u_m\cos t}\right)-\gamma_E\right]\sin^2 t.
\label{I1}
\end{equation}
Using here $\int_0^{\pi/2}\ln(\cos t)\sin^2 tdt=-(1+2\ln 2)\pi/8$ yields:
\begin{equation}
I=\frac{1}{2}\left(\ln\frac{4\tilde{L}_\theta}{u_m}-\gamma_E+\frac{1}{2}\right)\to\frac{1}{2}\ln\left(\frac{3.7\tilde{L}_\theta}{u_m}\right).
\end{equation}
Combining Eqs. (\ref{tam})-(\ref{I1}) gives Eqs. (\ref{tmm})-(\ref{Xii}).

\end{document}